\documentclass[aps,prd,floats,floatfix,nofootinbib,superscriptaddress]{revtex4-1}
\usepackage{graphicx}
\usepackage{amssymb}
\usepackage{epstopdf}
\usepackage{hyperref}

\usepackage{amsmath}
\usepackage{color}

\begin{document}

\title{Scalar decays to $\gamma\gamma$, $Z\gamma$, and $W\gamma$ in the Georgi-Machacek model}

\author{C\'eline Degrande}
\email{celine.degrande@cern.ch}
\affiliation{CERN, Theory Division, Geneva 23 CH-1211, Switzerland}

\author{Katy Hartling}
\email{kathryn.hartling@cnl.ca \\ Current address: {\it Canadian Nuclear Laboratories, 286 Plant Road, Chalk River, Ontario K0J 1J0, Canada}}
\author{Heather E.~Logan}
\email{logan@physics.carleton.ca} 

\affiliation{Ottawa-Carleton Institute for Physics, Carleton University, 1125 Colonel By Drive, Ottawa, Ontario K1S 5B6, Canada}

\date{August 29, 2017}                                  

\begin{abstract}
We compute the decay widths for the neutral and singly-charged Higgs bosons in the Georgi-Machacek model into the final states $\gamma\gamma$, $Z\gamma$, and $W\gamma$.  These decays are most phenomenologically interesting for the fermiophobic custodial fiveplet states $H_5^0$ and $H_5^{\pm}$ when their masses are below threshold for decays into $WW$, $ZZ$, or $WZ$.  We study the allowed branching ratios into these final states using scans over the allowed parameter space, and show how the model can be constrained by LEP searches for a fermiophobic Higgs boson decaying to two photons.
The calculation involves evaluating one-loop diagrams in which the loop contains particles with two different masses, some of which do not appear in the existing literature.  We give results for these diagrams in a form convenient for numerical implementation using the LoopTools package.

\end{abstract}

\maketitle 

\section{Introduction}

Since the discovery of a Standard Model (SM)-like Higgs boson at the CERN Large Hadron Collider (LHC)~\cite{Aad:2012tfa}, there has been considerable interest in models with extended Higgs sectors to be used as benchmarks for LHC searches for physics beyond the SM.  One such model is the Georgi-Machacek (GM) model~\cite{Georgi:1985nv,Chanowitz:1985ug}, which adds isospin-triplet scalar fields to the SM in a way that preserves custodial SU(2) symmetry.  This model is interesting because the isospin triplets can make a non-negligible contribution to electroweak symmetry breaking.  Its phenomenology has been studied extensively~\cite{Gunion:1989ci,Gunion:1990dt,HHG,Haber:1999zh,Aoki:2007ah,Godfrey:2010qb,Low:2010jp,Logan:2010en,Falkowski:2012vh,Chang:2012gn, Carmi:2012in, Chiang:2012cn,Chiang:2013rua,Kanemura:2013mc,Englert:2013zpa,Killick:2013mya,
Belanger:2013xza, Englert:2013wga,Efrati:2014uta,Hartling:2014zca,Chiang:2014hia,Chiang:2014bia,Godunov:2014waa,Hartling:2014aga,Chiang:2015amq,Chiang:2015rva,Arroyo-Urena:2016gjt,Chang:2017niy,Blasi:2017xmc,Zhang:2017och,Chiang:2017vvo}, and its parameter space has been constrained using the perturbativity and vacuum stability of the scalar potential~\cite{Aoki:2007ah,Chiang:2012cn,Hartling:2014zca}, the electroweak oblique parameters~\cite{Kanemura:2013mc,Englert:2013zpa,Chiang:2013rua,Hartling:2014aga}, $Z$-pole and $B$-physics observables~\cite{Haber:1999zh,Chiang:2012cn,Chiang:2013rua,Hartling:2014aga}, and direct collider searches~\cite{Chiang:2014bia,Khachatryan:2014sta,Aad:2015nfa,Chiang:2015kka,Logan:2015xpa,CMS:2016szz}.  The GM model has also been incorporated into Little Higgs~\cite{Chang:2003un,Chang:2003zn}, supersymmetric~\cite{Cort:2013foa,Garcia-Pepin:2014yfa,Delgado:2015bwa}, and neutrino seesaw~\cite{Godunov:2014waa} models. Extensions with an additional isospin doublet~\cite{Hedri:2013wea} and a singlet scalar dark matter candidate~\cite{Campbell:2016zbp} have also been considered, as have generalizations of the model to include higher-isospin scalars~\cite{Galison:1983qg,Robinett:1985ec,Logan:1999if,Chang:2012gn,Logan:2015xpa}. 

The most distinct phenomenological feature of the GM model is the presence of a custodial fiveplet of scalars, $(H_5^{++}, H_5^+, H_5^0, H_5^-, H_5^{--})$.  These scalars are fermiophobic and couple at tree level to $W$ or $Z$ boson pairs with a strength proportional to the isospin-triplet scalar fields' vacuum expectation value (vev).  Direct searches at the LHC for these custodial-fiveplet scalars have so far focused on scalar masses above 200~GeV~\cite{Khachatryan:2014sta,Aad:2015nfa,CMS:2016szz} (see also Refs.~\cite{Zaro:2015ika,deFlorian:2016spz}), where they decay predominantly into pairs of on-shell vector bosons.  For lower masses, the tree-level decays are forced off-shell and the loop-induced decays of $H_5^{\pm} \to W^{\pm} \gamma$ and $H_5^0 \to \gamma\gamma, Z\gamma$ can become important.  These final states offer sensitive new experimental probes.  The diphoton decay mode can also be used to take advantage of existing limits on the production of scalars decaying to photon pairs from the CERN Large Electron-Positron (LEP) collider~\cite{LEP2002} and the LHC~\cite{Delgado:2016arn}.

Our goal in this paper is to compute the loop-induced decay widths of the scalars in the GM model and study their behavior over the model's parameter space, focusing on scalar masses below 200~GeV.  This is made nontrivial by the fact that some diagrams appear in the decays $H_5^0 \to Z\gamma$ and $H_5^{\pm} \to W^{\pm} \gamma$ that have not previously been computed in the literature.  Some of these new diagrams also appear in the custodial-triplet scalar decay $H_3^{\pm} \to W^{\pm} \gamma$; we discuss this process for completeness although it is of less phenomenological interest because decays of $H_3^{\pm}$ to fermion pairs tend to dominate its branching ratios.

The challenge is diagrams in which the loop contains particles with two different masses.  Such ``heterogeneous'' loop diagrams are forbidden by gauge invariance in the familiar decays of the SM Higgs boson to two photons or two gluons; they are absent in the SM Higgs decay to $Z \gamma$ due to custodial symmetry.  Heterogeneous diagrams appear in two Higgs doublet models in the decay $H^{\pm} \to W^{\pm} \gamma$; these have been computed in Refs.~\cite{Arhrib:2006wd,Enberg:2013jba,Ilisie:2014hea}.\footnote{Heterogeneous diagrams contributing to neutral Higgs boson decays to $Z \gamma$ involving fermions and vector bosons in the loops have been computed in Refs.~\cite{Djouadi:1996yq} and \cite{Cai:2013kpa}, respectively.  These contributions do not appear in the GM model.}  In the two Higgs doublet model, the contributing diagrams involve top and bottom quarks, $H^{\pm}$ and a neutral scalar $h^0$ or $H^0$, and $W^{\pm}$ and a neutral scalar $h^0$ or $H^0$.  Explicit results for these loop diagrams have been given in Ref.~\cite{Ilisie:2014hea} as integrals over Feynman parameters.  For ease of numerical implementation, we recalculate them here in terms of the one-loop Passarino-Veltman integrals~\cite{Passarino:1978jh} in the notation used by the LoopTools package~\cite{Hahn:1998yk}.  Our results agree with those of Ref.~\cite{Ilisie:2014hea}.

The GM model admits additional heterogeneous diagrams not present in two Higgs doublet models.  These include diagrams that involve $W^{\pm}$ and $Z$, $Z$ and $H_5^{\pm}$, $W^{\pm}$ and $H_5^{\pm\pm}$, and $W^{\pm}$ and $H_5^{\pm}$.  These contribute to the decays $H_5^+ \to W^+ \gamma$, $H_3^+ \to W^+\gamma$, and $H_5^0 \to Z \gamma$.  By contributing to $H_5^0 \to Z \gamma$, the new diagrams can affect the branching ratio of $H_5^0 \to \gamma\gamma$ (though we find that the effect is numerically small).  We compute these new loop diagrams and give explicit results as integrals over Feynman parameters as well as in terms of the one-loop Passarino-Veltman integrals in the notation used by the LoopTools package.

With the new loop diagrams in hand, we implement the full one-loop decays $H_5^0 \to Z \gamma$, $H_5^{\pm} \to W^{\pm} \gamma$, and $H_3^{\pm} \to W^{\pm} \gamma$ into a private code based on GMCALC~1.2.0~\cite{Hartling:2014xma} (all other decays to $\gamma\gamma$ and $Z\gamma$ are already implemented in the public version of the code) and perform parameter scans to study the allowed range of branching ratios after imposing the theoretical and experimental constraints on the model.  We show that a large fraction of the parameter space with $H_5$ masses below about 110~GeV is excluded by LEP searches for fermiophobic Higgs production in $e^+e^- \to Z H_5^0$ with $H_5^0 \to \gamma\gamma$~\cite{LEP2002}.  Our results for the $H_5^0 \to \gamma \gamma$ branching ratio can also be combined with scalar pair-production cross sections to impose limits from LHC diphoton searches as in Ref.~\cite{Delgado:2016arn}; we leave this to future work.  These one-loop decays will be included in GMCALC~1.3.0 and higher.

This paper is organized as follows. In Sec.~\ref{sec:loopformulae} we present the results of the one-loop diagram calculations for the decays of the GM scalars to $V\gamma$. In Sec.~\ref{sec:decays} we assemble the familiar loop contributions with these new diagrams to compute the decay amplitudes for neutral scalars into $\gamma\gamma$ and $Z\gamma$ and for singly-charged scalars into $W\gamma$ in the GM model.  In Sec.~\ref{sec:BRs} we present numerical scans over the viable GM parameter space and apply the LEP limit on fermiophobic Higgs decays into two photons to constrain the model.  We summarize our conclusions in Sec.~\ref{sec:conclusions}.  For completeness, in Appendix~\ref{sec:GMmodel} we review the Lagrangian and physical spectrum of the GM model, in Appendix~\ref{app:couplings} we collect the Feynman rules for the GM model scalars that we use in this paper, and in Appendix~\ref{app:looptools} we summarize the LoopTools conventions for the one-loop Passarino-Veltman integrals used in our results.  Finally in Appendix~\ref{app:tHF} we give some details of the calculations in the 't~Hooft-Feynman gauge of processes involving Goldstone bosons or ghosts.

\section{One-loop diagrams for scalar decays to $V \gamma$}
\label{sec:loopformulae}

The decay amplitude for $H_i(k+q) \to V_{\nu}(k) \gamma_{\mu}(q)$ (where $V=\gamma,Z,W$) is forced by electromagnetic gauge invariance to take the form~\cite{Ilisie:2014hea}
\begin{equation}
	\mathcal{M} = \Gamma^{\mu \nu} \varepsilon^*_{\mu}(q) \varepsilon^*_{\nu}(k),
	\qquad {\rm with} \qquad
	\Gamma^{\mu\nu} = (g^{\mu\nu} k \cdot q - k^{\mu} q^{\nu}) S 
		+ i \epsilon^{\mu \nu\alpha\beta} k_{\alpha} q_{\beta} \tilde S,
	\label{eq:matrixelt}
\end{equation}
where $q$ and $k$ are the momenta and $\varepsilon_{\mu}(q)$ and $\varepsilon_{\nu}(k)$ are the polarization vectors of the photon and the gauge boson $V$, respectively.
The resulting decay partial width is
\begin{equation}
	\Gamma(H_i \to V \gamma) = \frac{m_{H_i}^3}{32 \pi \eta_V} 
		\left[ 1 - \frac{M_V^2}{m_{H_i}^2} \right]^3
		\left( |S|^2 + |\tilde S|^2 \right),
		\label{eq:decay}
\end{equation}
where $V = \gamma$, $Z$, or $W^+$.  Here $\eta_V$ is a symmetry factor that accounts for identical particles in the final state, with $\eta_{\gamma}=2$ and $\eta_Z = \eta_W = 1$.

\begin{figure}[t]
\begin{center}
\resizebox{0.32\textwidth}{!}{\includegraphics{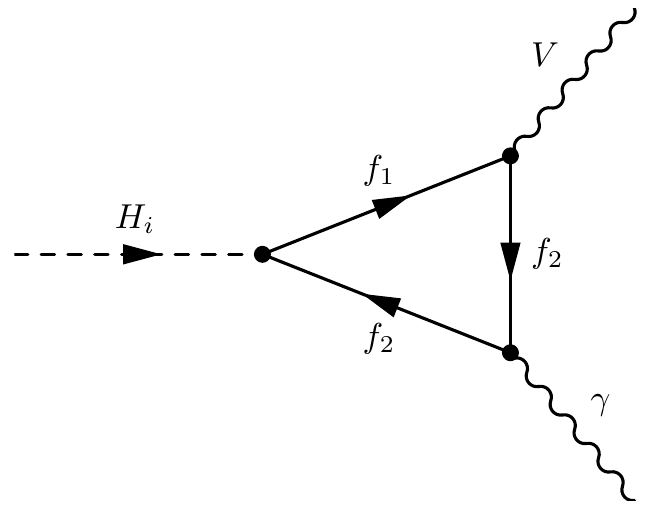}}
\resizebox{0.32\textwidth}{!}{\includegraphics{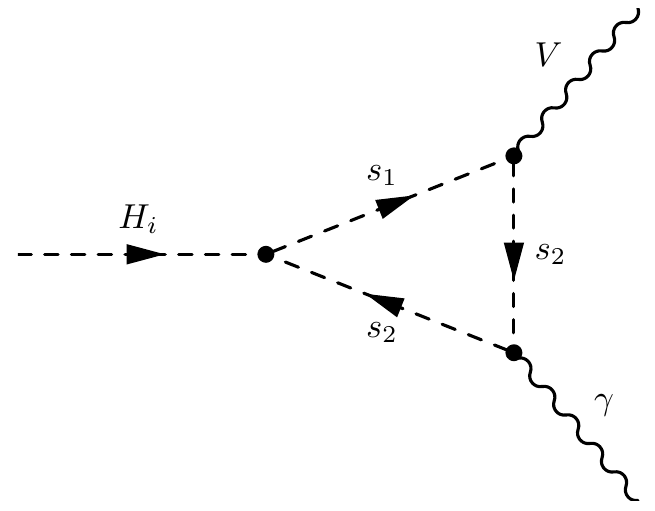}}
\resizebox{0.32\textwidth}{!}{\includegraphics{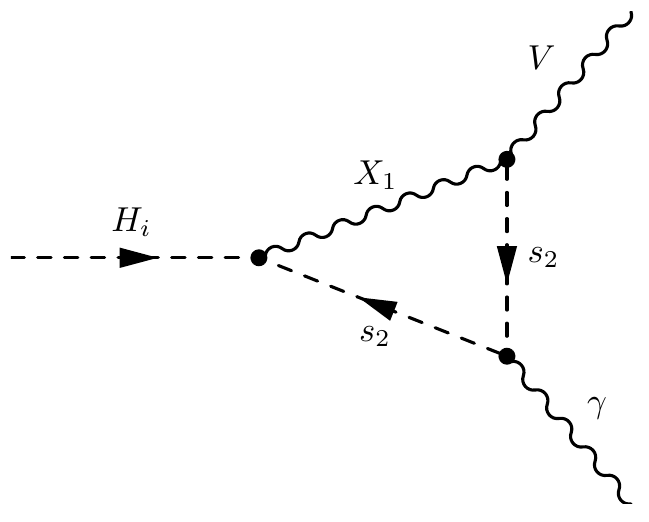}}
\resizebox{0.32\textwidth}{!}{\includegraphics{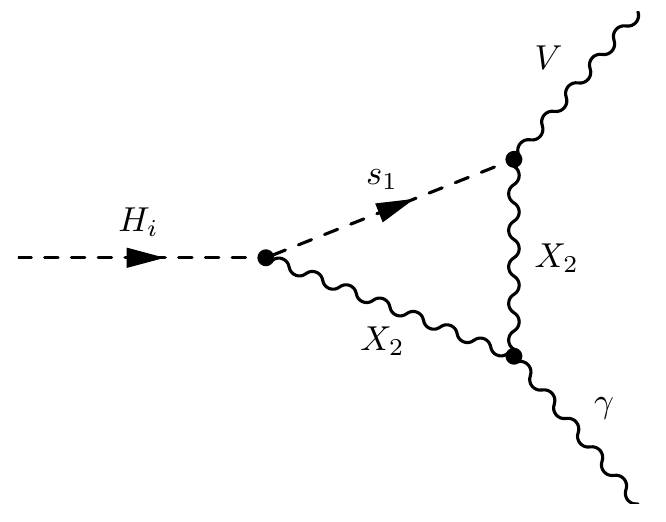}}
\resizebox{0.32\textwidth}{!}{\includegraphics{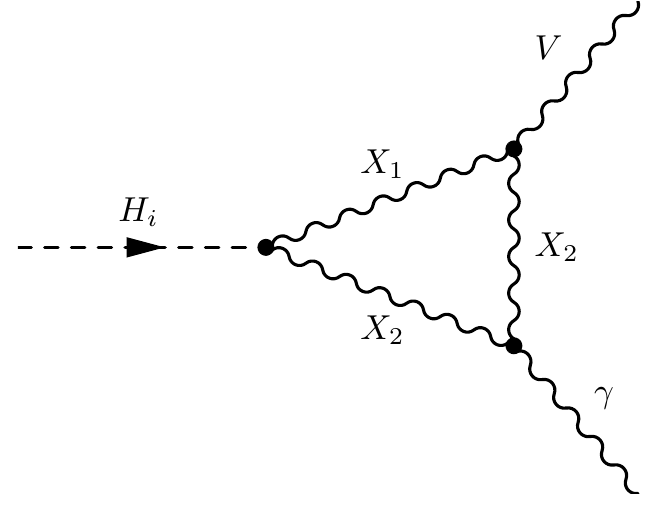}}
\resizebox{0.32\textwidth}{!}{\includegraphics{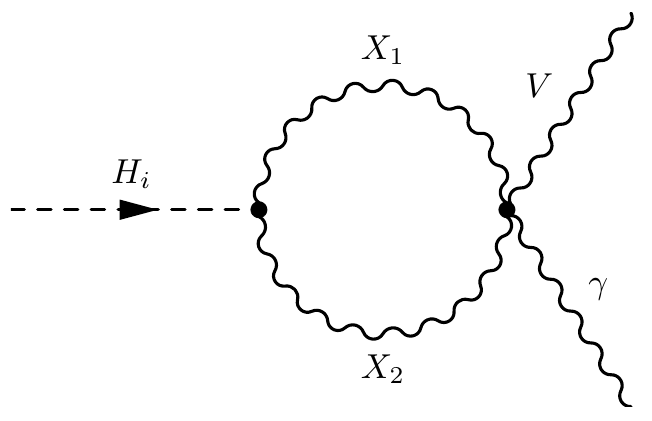}}
\caption{One-loop Feynman diagrams contributing to the process $H_i\rightarrow V\gamma$ in the GM model.  Here $V = \gamma$, $Z$ or $W^{\pm}$, $f_{1,2}$ denote fermions, $s_{1,2}$ denote scalars, and $X_{1,2}$ denote $W$ or $Z$ bosons.}
\label{fig:feyndiagrams}
\end{center}
\end{figure}

In calculating the scalar formfactor $S$, we follow the approach used by Ref.~\cite{Ilisie:2014hea} for the calculation of the one-loop amplitudes contributing to $H^+ \to W^+ \gamma$ in the Yukawa-aligned two Higgs doublet model (2HDM)~\cite{Pich:2009sp}. Ref.~\cite{Ilisie:2014hea} employed the clever strategy of computing only the coefficient of $k^{\mu} q^{\nu}$ in order to determine the form factor $S$. Neglecting all terms proportional to $g^{\mu\nu}$ significantly reduces the complexity of the calculations, as it reduces the number of Feynman diagrams that must be considered to those illustrated in Fig.~\ref{fig:feyndiagrams} and removes the need for renormalization.  The pseudoscalar formfactor $\tilde S$ receives contributions only from fermions in the loop as shown in the first diagram of Fig.~\ref{fig:feyndiagrams}.

To fix the signs of the charges of the particles appearing in the triangle diagrams of Fig.~\ref{fig:feyndiagrams}, we adopt the convention that $H_i$ is the incoming parent scalar and $V$ is an outgoing final-state vector boson.  The particle in the loop with subscript 1 propagates from the $H_i$ vertex to the $V$ vertex, while the particle in the loop with subscript 2 propagates from the $V$ vertex, through the photon vertex, back to the $H_i$ vertex.

The first diagram in Fig.~\ref{fig:feyndiagrams} has been computed in Ref.~\cite{Ilisie:2014hea}.  The second diagram has been computed in Ref.~\cite{Ilisie:2014hea} for the special case that $H_i$ and $s_2$ have the same mass.  The fourth diagram has been computed in Ref.~\cite{Ilisie:2014hea} for the special case that $X_2$ and $V$ have the same mass.  To our knowledge, the remaining diagrams have not appeared in the literature.  

In what follows we give our results for each diagram in the context of the GM model.  The results given in terms of integrals over Feynman parameters were computed in Unitarity gauge, while the results given in terms of LoopTools functions were computed in 't~Hooft-Feynman gauge including all relevant additional diagrams involving Goldstone bosons or ghosts.  We used dimensional regularization to handle divergences, which cancel in the final results.  The LoopTools conventions for the three-point integrals are summarized in Appendix~\ref{app:looptools}.  In each case we checked numerically that the two approaches agree to within the (percent-level) precision of our numerical integration over the Feynman parameters.  

The decays of the scalars $H_5^0 \to Z\gamma$, $H_3^0 \to Z\gamma$, and $H_5^+ \to W^+ \gamma$ have also been checked numerically using MadGraph5\_aMC@NLO~\cite{Alwall:2014hca} with the GM model renormalized by NLOCT~\cite{Degrande:2014vpa}.  It should be noted that these tools compute the full amplitude, including the coefficient of the $g^{\mu\nu}$ term in Eq.~(\ref{eq:matrixelt}), and so the electroweak renormalization of the model including tadpole and mixing counterterms is needed to obtain finite results.  Again, the numerical results agree at the percent level.  The decay $H_3^+ \to W^+ \gamma$ has not been checked using MadGraph5\_aMC@NLO because further development in the handling of contributions with on-shell cuts is still needed.

\subsection{Fermion loop diagram}
\label{sec:fff}

The first diagram in Fig.~\ref{fig:feyndiagrams} contributes to $H_3^+ \to W^+ \gamma$ with $f_1 = t$, $f_2 = b$, and with $f_1 = \bar b$, $f_2 = \bar t$.  The calculation is exactly as in Ref.~\cite{Ilisie:2014hea} with a translation from the Yukawa-aligned 2HDM coupling notation to the appropriate $\cot\theta_H$ dependence in the GM model (see Appendix~\ref{sec:GMmodel}).  The appropriate couplings are as given for the Type I 2HDM in Ref.~\cite{Ilisie:2014hea} with $\cot\beta \to \tan\theta_H$, i.e.,
\begin{equation}
	\varsigma_u = \tan\theta_H, \qquad \qquad
	\varsigma_d = \tan\theta_H.
\end{equation}
This yields the fermion loop contributions to the scalar and pseudoscalar formfactors as integrals over Feynman parameters $x$ and $z$~\cite{Ilisie:2014hea},
\begin{eqnarray}
	S_{H_3^+ W \gamma} \supset A_{ff^{\prime}}^{H_3^+W\gamma} &=& \frac{\alpha_{\rm em} N_c |V_{tb}|^2}{2 \pi v s_W} \tan\theta_H
		\int_0^1 dx \int_0^1 dz  \frac{I_f}
			{\Delta_{f}}, \label{eq:Af}
		\\
	\tilde S_{H_3^+ W \gamma} \supset \tilde A_{ff^{\prime}}^{H_3^+W\gamma} &=& \frac{\alpha_{\rm em} N_c |V_{tb}|^2}{2 \pi v s_W} \tan\theta_H
		\int_0^1 dx \int_0^1 dz  \frac{\tilde I_f}
			{\Delta_{f}}.\label{eq:tAf}
\end{eqnarray}
where $\alpha_{\rm em}$ is the electromagnetic fine structure constant, $N_c = 3$ is the number of colors, $v = (1/\sqrt{2} G_F)^{1/2} \simeq 246$~GeV is the SM Higgs vacuum expectation value (vev), $V_{tb}$ is the appropriate element of the Cabibbo-Kobayashi-Maskawa (CKM) quark-mixing matrix, and $s_W = \sin\theta_W$ is the sine of the weak mixing angle.  In the integrals we define
\begin{equation}
\Delta_{f} = M_W^2 x(x-1) + m_b^2 (1-x) + m_t^2 x + (M_W^2 - m_{3}^2) xz(1-x)\,,
\end{equation}
with $m_3$ being the mass of $H_3^+$, and 
\begin{align}
I_{f} & = [Q_t x + Q_b (1-x)] \left[-m_t^2 x(2xz-2z+1) + m_b^2 (1-x)(1-2xz)\right], \\
\tilde I_f & = [Q_t x + Q_b (1-x)][m_t^2 x + m_b^2 (1-x)].
\end{align}
The fermion electric charges are $Q_t = 2/3$ and $Q_b = -1/3$.

In terms of the LoopTools functions~\cite{Hahn:1998yk} (see Appendix~\ref{app:looptools} for conventions), the fermion loop contributions are given by
\begin{equation}
	A_{ff^{\prime}}^{H_3^+W\gamma} = A_{tbb}^{H_3^+ W \gamma} + A_{btt}^{H_3^+ W\gamma},
	\qquad
	\tilde A_{ff^{\prime}}^{H_3^+W\gamma} = \tilde A_{tbb}^{H_3^+ W \gamma} + \tilde A_{btt}^{H_3^+ W\gamma},
\end{equation}
where
\begin{eqnarray}
	A_{tbb}^{H_3^+ W \gamma} &=& \frac{\alpha_{\rm em} N_c |V_{tb}|^2}{2 \pi v s_W} \tan\theta_H
	Q_b \left[ m_t^2 \left( 2 C_{12} + 2 C_{22} + 3 C_2 + C_1 + C_0 \right) 
	\right. \nonumber \\ 
	&& \left. 
	- m_b^2 \left( 2 C_{12} + 2 C_{22} + C_2 - C_1 \right) \right] 
	(k^2, q^2, m_3^2; m_t^2, m_b^2, m_b^2),
		\\
	A_{btt}^{H_3^+ W \gamma} &=& \frac{\alpha_{\rm em} N_c |V_{tb}|^2}{2 \pi v s_W} \tan\theta_H
	Q_t \left[ -m_b^2 \left( 2 C_{12} + 2 C_{22} + 3 C_2 + C_1 + C_0 \right) 
	\right. \nonumber \\ 
	&& \left. 
	+ m_t^2 \left( 2 C_{12} + 2 C_{22} + C_2 - C_1 \right) \right] 
	(k^2, q^2, m_3^2; m_b^2, m_t^2, m_t^2),
		\\
	\tilde A_{tbb}^{H_3^+ W \gamma} &=& 
	\frac{\alpha_{\rm em} N_c |V_{tb}|^2}{2 \pi v s_W} \tan\theta_H
	Q_b \left[ -m_t^2 \left( C_1 + C_2 + C_0 \right) 
	\right. \nonumber \\
	&& \left.
	+ m_b^2 \left( C_1 + C_2 \right) \right] 
	(k^2, q^2, m_3^2; m_t^2, m_b^2, m_b^2),
		\\
	\tilde A_{btt}^{H_3^+ W \gamma} &=& 
	\frac{\alpha_{\rm em} N_c |V_{tb}|^2}{2 \pi v s_W} \tan\theta_H
	Q_t \left[ -m_b^2 \left( C_1 + C_2 + C_0 \right) 
	\right. \nonumber \\
	&& \left.
	+ m_t^2 \left( C_1 + C_2 \right) \right] 
	(k^2, q^2, m_3^2; m_b^2, m_t^2, m_t^2),
\end{eqnarray}
where $k^2 = M_W^2$ and $q^2 = 0$ are the final-state particles' invariant masses.

We can obtain a check of these formulas by artificially setting $m_b = m_t \equiv m_f$.  In this limit the $H^+ \bar t b$ coupling becomes purely pseudoscalar and the CP-even formfactor $A_{ff^{\prime}}^{H_3^+W\gamma}$ vanishes, while the CP-odd formfactor reduces to
\begin{eqnarray}
	\tilde A_{ff^{\prime}}^{H_3^+W\gamma} 
	&=& \frac{\alpha_{\rm em} N_c |V_{tb}|^2}{2 \pi v s_W} \tan\theta_H
	(Q_b + Q_t) \left[ -m_f^2 C_0 \right](k^2,q^2,m_3^2;m_f^2,m_f^2,m_f^2) \nonumber \\
	&=& \frac{\alpha_{\rm em} N_c |V_{tb}|^2}{2 \pi v s_W} \tan\theta_H
	(Q_b + Q_t) I_2(\tau_f,\lambda_f),
\end{eqnarray}
where in this case $\tau_f = 4 m_f^2/m_3^2$, $\lambda_f = 4 m_f^2/k^2$, and $I_2$ is the function that appears in the usual calculation of the fermion loop contribution to a CP-odd scalar decaying to $Z\gamma$~\cite{HHG} [see Eq.~(\ref{eq:I2})].  

\subsection{Scalar loop diagram}
\label{sec:sss}

The second diagram in Fig.~\ref{fig:feyndiagrams} contributes with two different scalar masses in the loop to $H_3^+ \to W^+ \gamma$.  (For $H_5^0 \to Z \gamma$ and $H_5^+ \to W^+ \gamma$ the scalar loop diagrams involve three scalars with the same masses.)  We define triple-scalar and vector-scalar-scalar couplings, with all particles incoming, in terms of the Feynman rules such that the triple-scalar vertex Feynman rule is $-i C_{H_i s_1^* s_2}$ and the vector-scalar-scalar vertex Feynman rule is $i e C_{V^* s_1 s_2^*}(p_1 - p_2)^{\mu}$, where $p_1$ and $p_2$ are the incoming momenta of $s_1$ and $s_2^*$, respectively, and $V^*$ is the incoming particle corresponding to outgoing vector boson $V$.  The photon coupling to two scalars is fixed by the Feynman rule $i e^2 Q_s (p_s - p_{s^*})^{\mu}$, where $p_s$ is the incoming momentum of the incoming scalar $s$, $p_{s^*}$ is the incoming momentum of outgoing scalar $s$ (or incoming $s^*$), and $Q_s$ is the electric charge in units of $e$ of the scalar $s$.  Explicit formulas for these couplings in the GM model are given in Appendix~\ref{app:couplings}.

The formfactor is given as an integral over Feynman parameters by
\begin{equation}
	S_{H_i V \gamma} \supset A_{s_1s_2s_2}^{H_i V\gamma} = - \frac{\alpha_{\rm em} Q_{s_2}}{\pi} C_{H_i s_1^* s_2} C_{V^* s_1 s_2^*} \int_0^1 dx \int_0^1 dz 
	\frac{I_{s_1s_2s_2}^{H_i V\gamma}}{\Delta_{s_1s_2s_2}},
	\label{eq:Asss}
\end{equation}
where 
\begin{eqnarray}
	\Delta_{s_1s_2s_2} &=& -M_V^2 x (1-x) + m_{s_1}^2 (1-x) + m_{s_2}^2 x + xz(1-x)(M_V^2 - m_{H_i}^2), \nonumber \\
	I_{s_1s_2s_2} &=& x^2 z (1-x)\,.
\end{eqnarray}
This agrees with the corresponding result of Ref.~\cite{Ilisie:2014hea} for the special case $m_{s_2} = m_{H_i}$.

In terms of the LoopTools functions, the scalar loop contribution is given by
\begin{equation}
	A_{s_1s_2s_2}^{H_i V\gamma} = - \frac{\alpha_{\rm em} Q_{s_2}}{\pi} 
	C_{H_i s_1^* s_2} C_{V^* s_1 s_2^*}
	\left[ C_{12} + C_{22} + C_2 \right] (k^2, q^2, m_{H_i}^2; m_{s_1}^2, m_{s_2}^2, m_{s_2}^2),
\label{eq:AsssLT}
\end{equation}
where $k^2 = M_V^2$ and $q^2 = 0$ are the final-state particles' invariant masses.

In the limit $m_{s_1} = m_{s_2} \equiv m_s$, this expression reduces to
\begin{equation}
	A_{s_1s_2s_2}^{H_i V\gamma} = - \frac{\alpha_{\rm em} Q_{s_2}}{\pi} 
	C_{H_i s_1^* s_2} C_{V^* s_1 s_2^*}
	\frac{1}{4 m_s^2} I_1(\tau_s, \lambda_s),
	\label{eq:s1eqs2}
\end{equation}
where in this case $\tau_s = 4 m_s^2/m_{H_i}^2$, $\lambda_s = 4 m_s^2/k^2$, and $I_1$ is the function that appears in the usual calculation of the scalar loop contribution to a CP-even scalar decaying to $Z \gamma$~\cite{HHG} [see Eq.~(\ref{eq:I1})].

\subsection{Vector-scalar-scalar loop diagram}
\label{sec:vss}

The third diagram in Fig.~\ref{fig:feyndiagrams} contributes to $H_3^+ \to W^+ \gamma$, $H_5^+ \to W^+ \gamma$, and $H_5^0 \to Z \gamma$.  For this diagram we need the scalar-vector-vector coupling, which is defined by the Feynman rule $i e^2 C_{s^* X V^*} g^{\mu\nu}$, again with all particles incoming.  Explicit expressions are given in Appendix~\ref{app:couplings}.

The formfactor is given as an integral over Feynman parameters by
\begin{equation}
	S_{H_i V \gamma} \supset A^{H_i V\gamma}_{Xss} 
	= -\frac{\alpha_{\rm em}^2\,Q_{s_2}\,C_{X_1^* H_i s_2} C_{s_2^*X_1V^*} }{M_{X_1}^2}\int_0^1 dx \int_0^1 dz \frac{I_{Xss}^{H_i V\gamma}}{\Delta_{Xss}}\,,
\label{eq:AssV}
\end{equation}
where 
\begin{eqnarray}
	\Delta_{Xss} &=& -x(1-x) M_V^2 +(1-x)M_{X_1}^2+ x m_{s_2}^2 
		- x z (1-x)(m_{H_i}^2-M_V^2), 
		\nonumber \\
	I_{Xss}^{H_i V\gamma} &=& x^2z\left[\frac{2}{3}x^2(1+z)+\frac{2}{3}x(1-2z)-1\right] m_{H_i}^2 	+x^2(z-1)\left[\frac{2}{3}x(2-x)(z-2)+1\right] M_V^2 \nonumber \\ 
	&& \quad
	+x\left[\frac{2}{3}(z+1)x^2-5x+4\right] M_{X_1}^2
	+x^2\left[-\frac{2}{3}(1+z)x+1 \right]m_{s_2}^2 .
\end{eqnarray}

In terms of the LoopTools functions, the vector-scalar-scalar loop contribution is given by
\begin{eqnarray}
	A^{H_i V\gamma}_{Xss} &=& - \alpha_{\rm em}^2\,Q_{s_2}\,C_{X_1^* H_i s_2} C_{s_2^*X_1V^*} 
	\left[ -2 (C_{12} + C_{22} + 2 C_1 + 3 C_2 + 2 C_0) \right. \nonumber \\
	&& \left.
	- 2 \left( \frac{m_{H_i}^2 - m_{s_2}^2}{M_{X_1}^2} \right) (C_{12} + C_{22} + C_2) \right]
	(k^2, q^2, m_{H_i}^2; M_{X_1}^2, m_{s_2}^2, m_{s_2}^2),
\label{eq:A_Xss}
\end{eqnarray}
where $k^2 = M_V^2$ and $q^2 = 0$ are the final-state particles' invariant masses.

\subsection{Scalar-vector-vector loop diagram}
\label{sec:svv}

The fourth diagram in Fig.~\ref{fig:feyndiagrams} contributes to $H_3^+ \to W^+ \gamma$, $H_5^+ \to W^+ \gamma$, and $H_5^0 \to Z \gamma$.
The formfactor is given as an integral over Feynman parameters by
\begin{equation}
	S_{H_i V \gamma} \supset A^{H_i V\gamma}_{sXX}  
	= \frac{\alpha_{\rm em}^2\,Q_{X_2}\, C_{X_2 H_i s_1^* } C_{s_1 X_2^* V^*} }{2 M_{X_2}^2}
		\int_0^1 dx \int_0^1 dz \frac{I_{sXX}^{H_i V\gamma}}{\Delta_{sXX}},
\label{eq:AsVV}
\end{equation}
where 
\begin{eqnarray}
	\Delta_{s XX} &=& -x(1-x) M_V^2 +(1-x)m_{s_1}^2+ x M_{X_2}^2 
	-x z (1-x)(m_{H_i }^2-M_V^2), \nonumber \\
I_{sXX}^{H_i\gamma V} &=& \left[\frac{2}{3}x^4(z-2)(z-1)+\frac{4}{3}x^3(2z^2-3z+1)+x^2(z-1)\right] M_V^2 \nonumber \\ 
&&  \quad
+\left[-\frac{2}{3}x^4z(1+z)+\frac{8}{3}x^3z(2-z)-3x^2z\right] m_{H_i}^2 +\left[-\frac{2}{3}x^3(1+4z)+x^2(6z-1)\right] m_{s_1}^2 \nonumber \\ 
&&  \quad
+\left[\frac{2}{3}x^3(4z+1)+3x^2(3-2z)\right] M_{X_2}^2.
\end{eqnarray}
This agrees with the corresponding result of Ref.~\cite{Ilisie:2014hea} for the special case $M_{X_2} = M_V$.\footnote{Note that our integrand $I_{sXX}$ is defined such that our integral differs from that in Ref.~\cite{Ilisie:2014hea} by a factor of 4.}

In terms of the LoopTools functions, the scalar-vector-vector loop contribution is given by
\begin{eqnarray}
	A^{H_i V\gamma}_{sXX}  &=& \alpha_{\rm em}^2\,Q_{X_2}\, C_{X_2 H_i s_1^* } C_{s_1 X_2^* V^*} 
	\left[ -2 C_{12} - 2 C_{22} + 4 C_1 + 2 C_2 \right. \nonumber \\
	&& \left.
	- 2 \left( \frac{m_{H_i}^2 - m_{s_1}^2}{M_{X_2}^2} \right) (C_{12} + C_{22} + C_2) \right]
	(k^2, q^2, m_{H_i}^2; m_{s_1}^2, M_{X_2}^2, M_{X_2}^2),
\label{eq:AsXX}
\end{eqnarray}
where $k^2 = M_V^2$ and $q^2 = 0$ are the final-state particles' invariant masses.

\subsection{Vector loop diagram}
\label{sec:vvv}

The fifth and sixth diagrams in Fig.~\ref{fig:feyndiagrams} contribute with two different gauge boson masses in the loop to $H_5^+ \to W^+ \gamma$.  The masses and couplings are given by $m_{H_i} = m_5$, $M_{X_1} = M_Z$, $M_{X_2} = M_V = M_W$, $Q_{X_2} = -1$, $C_{X_1 V^*X_2^*} = C_{ZW^-W^+} = c_W/s_W$, and $C_{H_i X_1^*x_2} = C_{H_5^+ Z W^-} = -v \sin\theta_H/2 c_W s_W^2$.

The formfactor is given as an integral over Feynman parameters by
\begin{equation}
	S_{H_i V \gamma} \supset \mathcal{A}_{X_1X_2X_2}^{H_i V\gamma} 
	= \frac{\alpha_{\rm em}^2}{2} Q_{X_2} C_{X_1 V^*X_2^*} C_{H_i X_1^*X_2}\int_0^1 dx
	\int_0^{1}dz\frac{I_{X_1X_2X_2}^{H_i V\gamma}}{\Delta_{X_1X_2X_2}},
\label{eq:AVVV}
\end{equation}
where 
\begin{eqnarray}
	\Delta_{X_1X_2X_2} &=& -M_V^2 x (1-x) + M_{X_1}^2 (1-x) + M_{X_2}^2 x  
	+ (M_V^2 - m_{H_i}^2) xz (1-x), 
		\nonumber \\
	I_{X_1X_2X_2}^{H_i V\gamma} &=& \frac{x}{3 M_{X_1}^2 M_{X_2}^2}\left[M_{X_1}^4\left(x\left[117-6x\left(34+16x\left[z-1\right]-29z\right)-72z\right]-12\right)\right. \nonumber \\
 && \left. +M_{X_2}^2\left(M_{X_2}^2x\left[9+6x\left(z-6-16x\left[z-1\right]\right)\right]+M_{X_1}^2\left[36+6x\left(x\left[40-18z\right] \right.\right.\right.\right. \nonumber \\
 && \left.\left.\left.\left.+32x^2\left[z-1\right]-3\left[3+4z\right]\right)\right]\right)+m_{H_i}^2x\left(m_{H_i}^2x\left[xz\left(2-24x\left[11+5x\left(z-1\right)-9z\right]z \right.\right.\right.\right.\nonumber \\
 && \left.\left.\left.\left. -90\left[z-2\right]z\right)-4z\left(1+9z\right)\right]+M_{X_2}^2\left[x\left(6z\left[13+2z\right]-2-12x\left[23+18x\left(z-1\right)\right.\right.\right.\right.\right. \nonumber \\
&& \left.\left.\left.\left.\left. -16z\right]z\right)\!-\!9z\right]+M_{X_1}^2\!\left[57z+x\left(2+12x\left[39+18x\left(z\!-\!1\right)-32z\right]z+6z\left[26z\!-\!51\right]\right)\right]\right) \right. \nonumber \\
&& \left.+M_V^2\left(M_{X_2}^2x\left[9z-12+x\left(64+12x\left[23+18x\left(z\!-\!1\right)-16z\right]\left(z\!-\!1\right)-2z\left[29+6z\right]\right)\right] \right.\right. \nonumber \\ 
 && \left.\left.+M_V^2x\left[3-3z-2x\left(z-1\right)^2\left(20+60x^3\left[z-1\right]-12x^2\left[9z-11\right]+x\left[45z-88\right]\right)\right] \right.\right. \nonumber \\
&& \left.\left.+M_{X_1}^2\,\left[x\left(72-57z+x\left[\left(454\!-\!156z\right)z-304-12x\left(39+18x\left[z\!-\!1\right]\!-\!32z\right)\!\left(z\!-\!1\right)\right]\right) \right.\right.\right. \nonumber\\
&& \left.\left.\left.-12\right]+m_{H_i}^2x\left[3z+x\left(2+z\left[76z-78\right]+4x\left[z-1\right]\left[1+\left(132x-60x^2-89\right)z \right.\right.\right.\right.\right. \nonumber \\
&& \left.\left.\left.\left.\left.+3\left(15-36x+20x^2\right)z^2\right]\right)\right]\right)\right].
\end{eqnarray}

The numerator can be equivalently expressed in the following, equally horrible, form,
\begin{eqnarray}
	I_{X_1X_2X_2}^{H_i V\gamma} &=& \frac{m_{H_i}^4}{M_{X_1}^2 M_{X_2}^2}\left[40 x^6 z^2 (1 - z) + 8z^2 x^5 (9z-11) +\frac{2}{3} x^4 z (1+90z-45z^2) - \frac{4}{3}x^3 z (1+9z) \right] 
\nonumber \\ 
&& + \frac{m_{H_i}^2 M_{V}^2}{M_{X_1}^2M_{X_2}^2}\left[80  x^6 z(z-1)^2  -16 x^5 z (z-1)(9z-11) +  \frac{4}{3} x^4 (z-1)\left(1-89z+45z^2\right) 
\right.\nonumber \\ 
&& \left. + \frac{2}{3}x^3 (z-1)(38z-1) + x^2 z\right] + \frac{m_{H_i}^2}{M_{X_1}^2} \left[72 x^5 z(1 - z) + 4 x^4 z (16z-23) + \frac{2}{3}x^3 \left(6z^2+39z-1\right) 
\right.\nonumber \\ 
&& \left. - 3 x^2 z\right] +   \frac{m_{H_i}^2}{M_{X_2}^2} \left[72 x^5 z (z-1) -4 x^4 z (32z-39) + \frac{2}{3} x^3 (1-153z+78z^2) + 19 x^2 z\right] 
\nonumber \\ 
&& + \frac{M_{V}^4}{M_{X_1}^2 M_{X_2}^2} \left[ 40 x^6 (1-z)^3 + 8 x^5 (z-1)^2 (9z-11) -\frac{2}{3} x^4 (z-1)^2 \left(45z-88\right)+ (1 - z) x^2
\right.\nonumber \\ 
&& \left. - \frac{40}{3} x^3(z-1)^2 \right] + \frac{M_{V}^2}{M_{X_1}^2} \left[ 72 x^5 (z-1)^2 - \frac{2}{3} x^3 \left(6z^2 + 29 z - 32\right)- 4 x^4 (z-1) (16z-23)
\right.\nonumber \\ 
&& \left. + x^2(3 z-4)\right] + \frac{M_{V}^2}{M_{X_2}^2} \left[ -72 x^5 (z-1)^2 + 4 x^4(z-1) (32z-39) + \frac{2}{3} x^3 \left(152-227z+78z^2\right)
\right.\nonumber \\ 
&& \left. +  x^2 (24-19 z) -   4 x\right] +  \frac{M_{X_2}^2}{M_{X_1}^2} \left[32 x^4 (1 - z) + 2x^3(z - 6) + 3 x^2\right] + 64 x^4 (z - 1) + 4 x^3 (20 - 9 z) 
\nonumber \\ 
&& -  6 x^2 (3 + 4 z)+ 12 x +\frac{M_{X_1}^2}{M_{X_2}^2} \left[ 32 x^4 (1 - z) + 2 x^3 (29 z - 34) - 3 x^2 (8z-13) - 4 x\right]\,.
\end{eqnarray}

In terms of the LoopTools functions, the vector loop contribution to $H_5^+ \to W^+ \gamma$ is given by
\begin{eqnarray}
	A_{ZWW}^{H_5^+ W \gamma} &=& - \frac{\alpha_{\rm em}}{2 \pi v} \sin\theta_H
	M_W M_Z \cot\theta_W
	\left[ \left( 12 C_{12} + 12 C_{22} + 12 C_2 + 6 C_0 \right) 
	\textcolor{white}{\frac{A}{B}}
	\right. \nonumber \\
	&& \left. + \frac{s_W^2}{c_W^2} \left( C_{12} + C_{22} + 2 C_1 + 3 C_2 + 2 C_0 \right)
	+ \frac{m_5^2}{M_W^2} \left( C_{12} + C_{22} + C_2 \right) \right]
	(k^2, q^2, m_5^2; M_Z^2, M_W^2, M_W^2),
\label{eq:AZWW}
\end{eqnarray}
where $k^2 = M_W^2$ and $q^2 = 0$ are the final-state particles' invariant masses.  In this case we have computed the specific diagrams appearing in $H_5^+ \to W^+ \gamma$ rather than a more generic case because we had to use the relations among the gauge, Goldstone, and ghost couplings to simplify the result of the 't Hooft-Feynman gauge calculation.  We note also that these simplifications make use of the final-state $W$ boson mass, so that this result is good for on-shell decays only.

\section{Scalar decays to $V \gamma$ in the GM model}
\label{sec:decays}

With the loop functions in hand, we now assemble all the contributions to compute the amplitudes for decays of the neutral scalars in the GM model into $\gamma\gamma$ and $Z\gamma$ and the singly-charged scalars into $W\gamma$.

\subsection{Decays to two photons}

In the GM model, the neutral scalars $h$, $H$, $H_3^0$, and $H_5^0$ can decay to two photons through the usual loop-induced processes.  Electromagnetic gauge invariance ensures that only a single particle runs around the loop in each diagram, so that the decay amplitudes $S$ and $\tilde S$ can be expressed in terms of the familiar functions given, e.g., in Ref.~\cite{HHG}.
The contributing particles in the loop are summarized in Table~\ref{tab:hgaga}.  Note that $h$, $H$, and $H_5^0$ are CP-even and hence decay via the formfactor $S$, while $H_3^0$ is CP-odd and decays via the formfactor $\tilde S$.  The CP-odd formfactor receives contributions only from fermions in the loop.  The state $H_5^0$ is fermiophobic and hence the fermion loop does not contribute to its decay.

\begin{table}
\begin{center}
\begin{tabular}{c c c c c}
\hline \hline
~$H_i$~ & Formfactor & ~$V$~ & ~$f$~ & $s$ \\
\hline\hline
$h$ & $S$ & $W^+$ & $t$ & $\{H_3^+, H_5^+, H_5^{++}\}$ \\
$H$ & $S$ & $W^+$ & $t$ & $\{H_3^+, H_5^+, H_5^{++}\}$ \\
$H_5^0$ & $S$ & $W^+$ & -- & $\{H_3^+, H_5^+, H_5^{++}\}$ \\
$H_3^0$ & $\tilde S$ & -- & $t$ & -- \\
\hline \hline
\end{tabular}
\caption{Particles in the loop that contribute to the decay $H_i \to \gamma\gamma$.  For the fermion contributions we include only the dominant top quark contribution.}
\label{tab:hgaga}
\end{center}
\end{table}

For the CP-even scalars $h$, $H$, and $H_5^0$, the decay is described by Eq.~(\ref{eq:decay}) with $\tilde{S}=0$ and~\cite{HHG}~\footnote{Note that $1/2 \pi v = g/4 \pi M_W$.}
\begin{equation}
	S_{H_i\gamma\gamma} =  \frac{\alpha_{\rm em}}{2 \pi v}
	\left[\beta_W^{H_i} F_1(\tau_W) + \sum_f N_{cf} Q_f^2 \beta_f^{H_i} F_{1/2}(\tau_f) 
	+ \sum_s \beta_s^{H_i} Q_s^2 F_0(\tau_s) \right], \label{eq:hH2gam}
\end{equation}
where $N_{cf} = 3$ for quarks and 1 for leptons, $Q_j$ is the electric charge of particle $j$ in units of $e$, and the sums run over all fermions and scalars that can propagate in the loop for the parent scalar $H_i$. In practice the charged scalars are $s=\{H_3^+,H_5^+,H_5^{++}\}$ and we keep only the top quark contribution to the fermion loop, $f=t$. The coupling coefficients $\beta_{j}^{H_i}$ are defined as,
\begin{equation}
	\beta_W^{H_i} = \frac{C_{H_i W^+W^-} e^2 }{g M_W}, \quad\quad 
	\beta_f^{H_i} = -\frac{C_{H_i f\bar{f}} v}{m_f}, \quad\quad 
	\beta_s^{H_i} = \frac{C_{H_i ss^*} v}{2 m_s^2},
	\label{eq:beta}
\end{equation}
for a propagating $W$ boson, fermion $f$, and scalar $s$, respectively. The couplings $C_{ijk}$ are given in Appendix~\ref{app:couplings}.  In the case of the $W$ boson and fermion loops, these factors $\beta_{W,f}^{H_i}$ are equal to the usual ratios $\kappa_{W,f}^{H_i}$ of the scalar coupling to $WW$ or $f\bar f$ normalized to the corresponding SM Higgs coupling as described in Ref.~\cite{LHCHiggsCrossSectionWorkingGroup:2012nn}.  Note that $\beta_f^{H_5^0} = 0$ because the $H_5$ states are fermiophobic.

The loop factors are given in terms of the usual functions for particles of spin $0$, $1/2$ and $1$~\cite{HHG},
\begin{eqnarray}
	F_1(\tau_W) &=& 2+3\tau_W+3\tau_W(2-\tau_W) f(\tau_W), \nonumber \label{F1}\\
	F_{1/2}(\tau_f) &=& -2\tau_f[1+(1-\tau_f) f(\tau_f)], \nonumber \label{eq:F12} \\
	F_0(\tau_s) &=& \tau_s[1-\tau_s f(\tau_s)],\label{F0}
\end{eqnarray}
where $\tau_j = 4 m_j^2/m_{H_i}^2$ and
\begin{equation}
	f(\tau) = \left\{ \begin{array}{l l}
	\left[\sin^{-1} \left(\sqrt{\frac{1}{\tau}}\right) \right]^2 & \quad  {\rm if} \ \tau \geq 1, \\
	-\frac{1}{4}\left[ \log \left(\frac{\eta_+}{\eta_-}\right) - i \pi \right]^2 & \quad  {\rm if} \ \tau < 1, \\
	\end{array} \right.
	\label{feq}
\end{equation}
with $\eta_{\pm} = 1 \pm \sqrt{1-\tau}$.

For the CP-odd scalar $H_3^0$, the decay is described by Eq.~(\ref{eq:decay}) with $S = 0$ and~\cite{HHG}
\begin{equation}
	\tilde{S}_{H_3^0 \gamma\gamma} = \frac{\alpha_{\rm em}}{2 \pi v}
	\left[ \sum_f N_{cf} Q_f^2 \beta_f^{H_3^0} F^A_{1/2}(\tau_f) \right],
	\label{eq:H32gam}
\end{equation}
where $F_{1/2}^A = -2\tau_f f(\tau_f)$ and $\beta_f^{H_3^0}$ is defined as in Eq.~(\ref{eq:beta}).  Again we will include only the top quark in the loop, $f = t$.

\subsection{Decays to $Z\gamma$}

The neutral scalars $h$, $H$, $H_3^0$, and $H_5^0$ can also decay to $Z\gamma$ through a loop.  For this decay, loops involving particles with two different masses can appear, because the $Z$ boson (unlike the photon) can couple to two different-mass particles.  These new diagrams arise only in the decay of the custodial-fiveplet scalar $H_5^0 \to Z \gamma$, because custodial symmetry is enough to forbid them in the decays of the custodial-singlet scalars $h$ and $H$, and the CP-odd scalar $H_3^0 \to Z\gamma$ decay receives contributions only from loops of fermions, whose couplings to the $Z$ boson are flavor-diagonal.

As before, the CP-even scalars $h$, $H$, and $H_5^0$ decay via the formfactor $S$, while the CP-odd scalar $H_3^0$ decays via the formfactor $\tilde S$.  The state $H_5^0$ is fermiophobic and hence the fermion loop does not contribute to its decay.  The contributing particles in the loop are summarized in Table~\ref{tab:hgZ}.

\begin{table}
\begin{tabular}{c c c c c c c}
\hline \hline
~$H_i$~ & Formfactor & ~$V$~ & ~$f$~ & $s$ & $Xss$ & $sXX$ \\
\hline\hline
$h$ & $S$ & $W^+$ & $t$ & $\{H_3^+, H_5^+, H_5^{++}\}$ & -- & -- \\
$H$ & $S$ & $W^+$ & $t$ & $\{H_3^+, H_5^+, H_5^{++}\}$ & -- & -- \\
$H_5^0$ & $S$ & $W^+$ & -- & $\{H_3^+, H_5^+, H_5^{++}\}$ & $\{W^+ H_5^+ H_5^+, W^- H_5^- H_5^-\}$ & $\{H_5^+ W^+ W^+, H_5^- W^- W^-\}$ \\
$H_3^0$ & $\tilde S$ & -- & $t$ & -- & -- & -- \\
\hline\hline
\end{tabular}
\caption{Particles in the loop that contribute to the decay $H_i \to Z \gamma$.  For the fermion contributions we include only the dominant top quark contribution.  The $H_5^0 \to Z \gamma$ decay receives contributions from vector-scalar-scalar diagrams as computed in Sec.~\ref{sec:vss} and from scalar-vector-vector diagrams as computed in Sec.~\ref{sec:svv}.}
\label{tab:hgZ}
\end{table}

For the custodial-singlet CP-even scalars $h$ and $H$, the decay is described by Eq.~(\ref{eq:decay}) with $\tilde S = 0$ and
\begin{equation}
	S_{H_i Z\gamma} = -\frac{\alpha_{\rm em}}{2 \pi v}
	\left[\beta_W^{H_i} A_W^{H_iZ\gamma} + \sum_f \beta_f^{H_i} A_f^{H_iZ\gamma} 
	+ \sum_s \beta_s^{H_i} A_s^{H_iZ\gamma}\right],
\end{equation}
where the sums run over all fermions and scalars that can propagate in the loop for the parent scalar $H_i$.  In practice the contributing scalars are $s = \{ H_3^+, H_5^+, H_5^{++} \}$ and we keep only the top quark contribution to the fermion loop, $f = t$.  
The coupling coefficients $\beta_j^{H_i}$ are the same as in Eq.~(\ref{eq:beta}).
The loop factors are given as usual by~\cite{HHG}\footnote{Here we have rewritten $A_W^{H_i Z \gamma}$ in a form that clearly separates the kinematic dependence of the loop diagrams on $M_W$ and $M_Z$ (encoded in $\lambda_W$ and $\tau_W$) from their dependence on the triple- and quartic-gauge couplings, following Appendix B of Ref.~\cite{Hartling:2014zca}.}
\begin{eqnarray}
	A_W^{H_iZ\gamma} &=& - \frac{c_W}{s_W} \left\{ 
		\left[ 8 - \frac{16}{\lambda_W} \right] I_2(\tau_W,\lambda_W)
		+ \left[ \frac{4}{\lambda_W} \left( 1 + \frac{2}{\tau_W} \right) 
			- \left( 6 + \frac{4}{\tau_W} \right) \right] I_1(\tau_W,\lambda_W) \right\},
		\label{eq:AWgamZ} \\
	A_f^{H_iZ\gamma} &=& N_{cf} \frac{-2 Q_f \left(T^{3L}_f - 2 Q_f s^2_W \right)}{ s_W c_W}\left[ I_1(\tau_f,\lambda_f)-I_2(\tau_f,\lambda_f) \right],\label{eq:AFgamZ}\\
	A_s^{H_iZ\gamma} &=& 2 C_{Z ss^*}\,Q_s \,I_1\left(\tau_s,\lambda_s\right)\,, 
	\label{eq:ASgamZ}
\end{eqnarray}
where $T_f^{3L}$ is the third component of isospin of the left-handed fermion $f$ ($T_f^{3L} = 1/2$ for the top quark), $\tau_j = 4 m_j^2/m_{H_i}^2$, $\lambda_j = 4 m_j^2/M_Z^2$, and the loop functions are
\begin{eqnarray}
	 I_1(a,b) &=& \frac{ab}{2(a-b)} + \frac{a^2b^2}{2(a-b)^2} \left[f(a) - f(b)\right]
	 + \frac{a^2b}{(a-b)^2} \left[g(a) - g(b)\right], \label{eq:I1} \\
	 I_2(a,b) &=& -\frac{ab}{2(a-b)} \left[f(a) - f(b)\right]. 
	 \label{eq:I2}
\end{eqnarray}
The function $f(\tau)$ was given in Eq.~(\ref{feq}), and
\begin{equation}
	g(\tau) = \left\{ \begin{array}{l l}
	\sqrt{\tau-1} \sin^{-1} \left(\sqrt{\frac{1}{\tau}}\right) & \quad  {\rm if} \ \tau \geq 1, \\
	\frac{1}{2} \sqrt{1-\tau} \left[ \log \left(\frac{\eta_+}{\eta_-}\right) - i \pi \right] 
		& \quad  {\rm if} \ \tau < 1,
	\end{array} \right.
	\label{geq}
\end{equation}
with $\eta_{\pm}$ defined as for $f(\tau)$. 
The couplings $C_{Z s s^*}$ are given in Appendix~\ref{app:couplings}. 

For the custodial-fiveplet CP-even scalar $H_5^0$, the decay is described by Eq.~(\ref{eq:decay}) with $\tilde S = 0$ and
\begin{eqnarray}
	S_{H_5^0Z\gamma} &=& -\frac{\alpha_{\rm em}}{2 \pi v}
	\left[\beta_W^{H_5^0} A_W^{H_5^0Z\gamma} 
	+ \sum_s \beta_s^{H_5^0} A_s^{H_5^0Z\gamma}\right] \nonumber \\
	&& + A_{W^+H_5^+H_5^+}^{H_5^0Z\gamma} + A_{W^-H_5^-H_5^-}^{H_5^0Z\gamma} 
	+ A_{H_5^+W^+W^+}^{H_5^0Z\gamma} + A_{H_5^-W^-W^-}^{H_5^0Z\gamma},
	\label{eq:Ahgz}
\end{eqnarray}
where the sum over scalars runs over $s = \{H_3^+, H_5^+, H_5^{++}\}$ as before and $\beta_{W,s}^{H_5^0}$ are defined as in Eq.~(\ref{eq:beta}). The novel contributions are the last four terms, which come from the vector-scalar-scalar loop (Sec.~\ref{sec:vss}) and the scalar-vector-vector loop (Sec.~\ref{sec:svv}) involving a $W$ boson and an $H_5^{\pm}$.  Our conventions for these diagrams are such that the two directions of electric charge flow must be included explicitly, but this is simplified by the fact that $A_{W^+H_5^+H_5^+}^{H_5^0Z\gamma} = A_{W^-H_5^-H_5^-}^{H_5^0Z\gamma}$ and $A_{H_5^+W^+W^+}^{H_5^0Z\gamma} = A_{H_5^-W^-W^-}^{H_5^0Z\gamma}$.
There are no fermion loop contributions because $H_5^0$ is fermiophobic.

For the CP-odd scalar $H_3^0$, the decay is described by Eq.~(\ref{eq:decay}) with $S = 0$ and 
\begin{equation}
	\tilde{S}_{H_3^0 Z\gamma} = -\frac{\alpha_{\rm em}}{2 \pi v}
	\left[\sum_f \beta_f^{H_3^0} \tilde A_f^{H_3^0Z\gamma}\right],
\end{equation} 
where the CP-odd fermion loop function is~\cite{HHG}
\begin{equation}
	\tilde A_f^{H_iZ\gamma} = N_{cf} \frac{-2\,Q_f\left(T^{3L}_f-2\,Q_f s_W^2 \right)}{ s_W c_W }\left[-I_2(\tau_f,\lambda_f)\right]\,,\label{eq:ZgamAA}\\
\end{equation}
with $I_2$ given by Eq.~(\ref{eq:I2}) and $\beta_f^{H_3^0}$ defined as in Eq.~(\ref{eq:beta}).  Again we will include only the top quark in the loop, $f = t$.

\subsection{Decays to $W^+\gamma$}

The singly-charged scalars $H_3^+$ and $H_5^+$ of the GM model can decay to $W^+ \gamma$ through a loop.  These states are not CP eigenstates and hence their decays generically receive contributions from both $S$ and $\tilde S$ in Eq.~(\ref{eq:decay}).  Indeed, both $S$ and $\tilde S$ contribute to $H_3^+ \to W^+ \gamma$.  However, because $\tilde S$ is generated only by fermions in the loop, the decay of the fermiophobic scalar $H_5^+ \to W^+ \gamma$ receives contributions only from $S$.

For $H_3^+ \to W^+ \gamma$, the decay is described by Eq.~(\ref{eq:decay}) with 
\begin{eqnarray}
	S_{H_3^+W\gamma} &=& \sum_f A^{H_3^+W\gamma}_{ff^{\prime}}
	+ \sum_{s_1,s_2} A^{H_3^+W\gamma}_{s_1s_2s_2} 
	+ \sum_{X,s} A^{H_3^+W\gamma}_{Xss} 
	+ \sum_{s,X} A^{H_3^+W\gamma}_{sXX}, \\
 	\tilde{S}_{H_3^+W\gamma} &=& \sum_f \tilde A_{ff^{\prime}}^{H_3^+W\gamma}.
\end{eqnarray} 
The particles that contribute to the sums are summarized in Table~\ref{tab:H3ppart}.  This calculation requires the fermion diagram of Sec.~\ref{sec:fff}, the scalar diagram of Sec.~\ref{sec:sss}, the vector-scalar-scalar diagram of Sec.~\ref{sec:vss}, and the scalar-vector-vector diagram of Sec.~\ref{sec:svv}.

\begin{table}
\begin{tabular}{c c c}
\hline \hline
Diagram & Formfactor & Particles \\
\hline
$f_1 f_2 f_2$ & $S, \tilde S$ & $tbb, \bar b \bar b \bar t$ \\
$s_1s_2s_2$ & $S$ & $hH_3^-H_3^-, HH_3^-H_3^-, H_5^0H_3^-H_3^-, H_3^0H_5^-H_5^-, H_5^{++}H_3^+H_3^+, H_3^-H_5^{--}H_5^{--}$\\
${Xss}$ & $S$ & $ZH_5^-H_5^-, W^-H_5^{--}H_5^{--}$\\
${s XX}$ & $S$ & $hW^-W^-, HW^-W^-, H_5^0W^-W^-, H_5^{++}W^+W^+$\\
\hline\hline
\end{tabular}
\caption{Particles in the loop that contribute to the decay $H_3^+ \to W^+ \gamma$.  For the fermion contribution we include only the third-generation quark loops.}
\label{tab:H3ppart}
\end{table}

For $H_5^+ \to W^+ \gamma$, the decay is described by Eq.~(\ref{eq:decay}) with $\tilde S = 0$ and
\begin{equation}
	S_{H_5^+W\gamma} = \sum_{s_1,s_2} A^{H_5^+W\gamma}_{s_1s_2s_2} 
	+ \sum_{X,s} A^{H_5^+W\gamma}_{Xss} 
	+ \sum_{s,X} A^{H_5^+W\gamma}_{sXX} 
	+ {A}_{ZWW}^{H_5^+W\gamma} \,.
\end{equation}
The particles that contribute to the sums are summarized in Table~\ref{tab:H5ppart}.
This calculation requires the scalar diagram of Sec.~\ref{sec:sss}, the vector-scalar-scalar diagram of Sec.~\ref{sec:vss}, the scalar-vector-vector diagram of Sec.~\ref{sec:svv}, and the vector diagrams of Sec.~\ref{sec:vvv}.  Note that the scalars $s_1$ and $s_2$ that run in the loop for $A^{H_5^+W\gamma}_{s_1s_2s_2}$ always have the same mass as each other due to the custodial symmetry, so that the scalar loop integral reduces in this case to the familiar $Z\gamma$ loop function as in Eq.~(\ref{eq:s1eqs2}).

\begin{table}
\begin{tabular}{c c c}
\hline \hline
Diagram & Formfactor & Particles \\
\hline
${s_1s_2s_2}$ & $S$ & $H_3^0H_3^-H_3^-, H_5^0H_5^-H_5^-, H_5^-H_5^{--}H_5^{--}, H_5^{++}H_5^{+}H_5^+$\\
${Xss}$ & $S$ & $ZH_5^-H_5^-, W^-H_5^{--}H_5^{--}$\\
${sXX}$ & $S$ & $H_5^0 W^-W^-, H_5^{++} W^+W^+$\\
${X_1X_2X_2}$ & $S$ & $Z W^-W^-$ \\
\hline\hline
\end{tabular}
\caption{Particles in the loop that contribute to the decay $H_5^+ \to W^+ \gamma$.}
\label{tab:H5ppart}
\end{table}

\subsection{Competing decay modes}

In order to compute the branching ratios for the loop-induced decays, we need the partial widths for all competing decay modes of the neutral and singly-charged scalars.  For this we use the decay partial width calculations for the scalars of the GM model as implemented in GMCALC~1.2.0~\cite{Hartling:2014xma}, which includes the following processes:
\begin{enumerate}
\item Tree-level decays to $V_1V_2$, with $V = W$ or $Z$, including full doubly-offshell effects;
\item Tree-level decays to one scalar and one vector boson, using the two-body expression when kinematically allowed and taking the vector boson off-shell otherwise;
\item Tree-level decays to two scalars (two-body only);
\item Decays to gluon pairs, including partial QCD corrections at the level of Ref.~\cite{Djouadi:1995gt};
\item Decays to fermion pairs (two-body only), including partial QCD corrections at the level of Ref.~\cite{Djouadi:1995gt}.
\end{enumerate}

In our numerical analysis we will be most interested in $H_5$ masses below the $VV$ threshold, where the loop-induced decays can obtain large branching ratios.  For such $H_5$ masses our inclusion of the doubly-offshell effects in the competing decays $H_5^0 \to WW, ZZ$ and $H_5^+ \to W^+Z$ is essential.  Also interesting are $H_3^+$ decays to $W^+ \gamma$ below the threshold for $H_3^+ \to t \bar b$.  We have not included off-shell top quark effects in this competing decay; we leave this improvement to future work.

For very light charged (neutral) scalars below the $W$ ($Z$) boson mass, off-shell loop-induced decays to $W^*\gamma$ ($Z^*\gamma$) can become relevant.  These could in principle be implemented by taking the $W$ ($Z$) boson off-shell with a Breit-Wigner distribution as in Ref.~\cite{Romao:1998sr}.  However, this approach neglects non-resonant box diagram contributions to the full $H_i \to f \bar f \gamma$ process (so-called Dalitz decays~\cite{Abbasabadi:1996ze,Sun:2013rqa,Passarino:2013nka}).  Furthermore, our result for the vector loop contribution to $H_5^+ \to W^+ \gamma$ in terms of the LoopTools functions is valid for an on-shell final-state $W$ boson only.  For these reasons, in our numerical implementation we compute the scalar decays to $W\gamma$ ($Z\gamma$) as strictly two-body decays.  
This will result in a counterintuitive resurgence of the off-shell $H_5^+ \to W^+ Z$ branching ratio for $m_5 < M_W$, when the two-body $H_5^+ \to W^+ \gamma$ decay is forbidden in our calculation.  Our numerical results for $H_5^+$ decay branching ratios are therefore not to be trusted for $m_5 < M_W$.  For these masses, one can rely instead on searches for $H_5^{++} \to W^+ W^+$, for which there is no competing decay, and for $H_5^0 \to \gamma\gamma$, which will dominate over any off-shell $H_5^0 \to Z^* \gamma$ contribution.  In our numerical analysis in the next section we consider only $H_5^+$ masses above $M_W$ and $H_5^0$ masses above $M_Z$.

\section{Numerical results} 
\label{sec:BRs}

In this section we study the branching ratios of the scalar decays to $\gamma\gamma$, $Z\gamma$, and $W\gamma$ in the GM model.  We set $m_h = 125$~GeV and scan over the full allowed parameter space of the model using a modified version of GMCALC~1.2.0~\cite{Hartling:2014xma} into which we have implemented the new one-loop decays.

GMCALC~1.2.0 lets us impose the theoretical constraints from perturbative unitarity of the quartic couplings in the Higgs potential and the stability of the correct electroweak vacuum~\cite{Hartling:2014zca}, as well as indirect constraints from $b\rightarrow s \gamma$~\cite{Hartling:2014aga}.  We also impose an upper bound on $\sin\theta_H$ as a function of $m_5$ determined in Ref.~\cite{Chiang:2014bia} from an ATLAS measurement of the like-sign $WW$ cross section in vector boson fusion at the 8~TeV LHC~\cite{Aad:2014zda}, which would be increased by $H_5^{++}$ production.\footnote{Other LHC searches for vector boson fusion production of $H_5^{++}$~\cite{Khachatryan:2014sta} or $H_5^+$~\cite{Aad:2015nfa,CMS:2016szz} consider only masses above 200~GeV.}  Finally we require that $m_3, m_5 \geq 76$~GeV; the lower bound on $m_5$ was found in Ref.~\cite{Logan:2015xpa} based on an ATLAS search for anomalous like-sign dimuon production at 8~TeV~\cite{ATLAS:2014kca,Kanemura:2014ipa}, and the lower bound on $m_3$ comes from the LEP search for charged Higgs pair production in the Type I two Higgs doublet model~\cite{Searches:2001ac}, where we require $m_3 \leq m_5$ to prevent off-shell decays $H_3^+ \to H_5^+ Z, H_5^0 W^+$.  In our scans we require $76~{\rm GeV} \leq m_5 \leq 200$~GeV and scan all other parameters over their allowed ranges.  The theoretical constraints force $m_3 \lesssim 600$~GeV in these scans.

\begin{figure}
\begin{center}
\resizebox{0.49\textwidth}{!}{\includegraphics{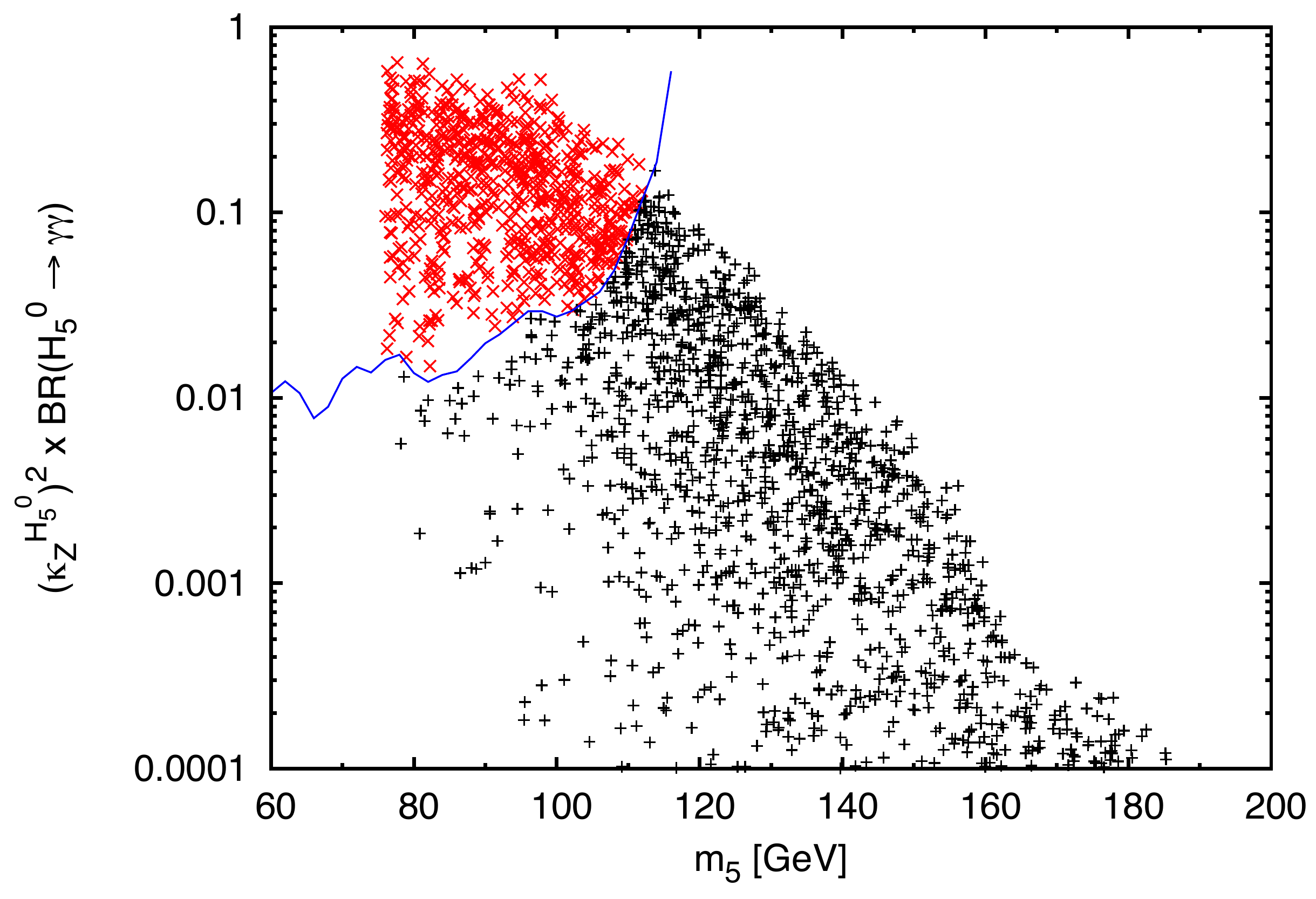}}
\resizebox{0.49\textwidth}{!}{\includegraphics{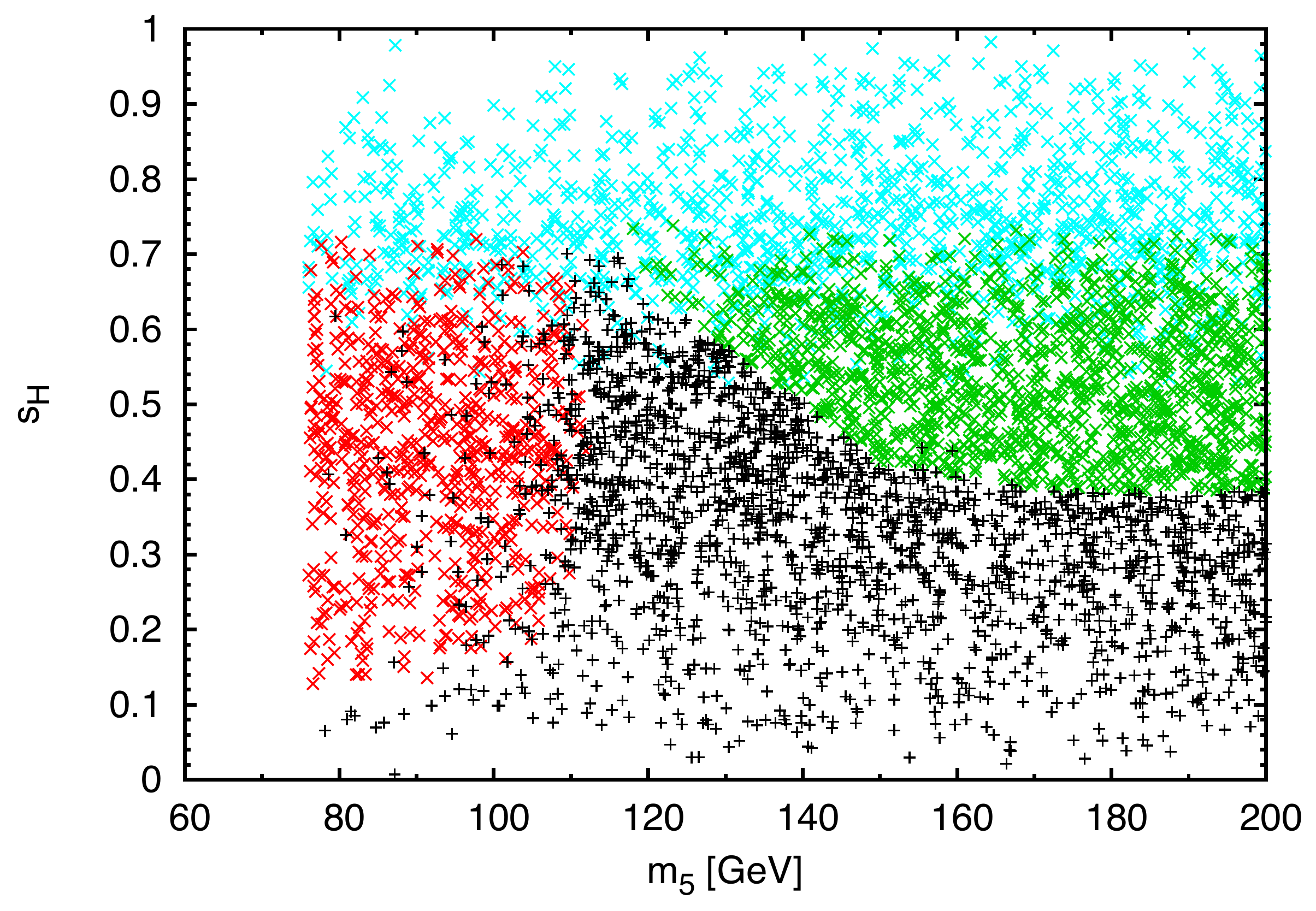}}
\caption{Left: BR$(H_5^0 \to \gamma\gamma)$ multiplied by the square of the $H_5^0ZZ$ coupling as a function of $m_5$, showing the region excluded by a LEP search for $e^+e^- \to ZH$ with $H \to \gamma\gamma$ (blue line).  Red points are excluded by the LEP search and black points are allowed.  All other experimental constraints are satisfied. Right: Scan points in the $m_5$--$s_H$ plane. All points shown satisfy the theory constraints and the lower bounds $m_5, m_3 \geq 76$~GeV.  Cyan points are excluded by $b \to s \gamma$, green points by the ATLAS like-sign $WWjj$ bound, and red points by the LEP constraint.  Black points are allowed.}
\label{fig:LEP}
\end{center}
\end{figure}

Our new result for $\Gamma(H_5^0 \to Z \gamma)$ allows us to make an accurate calculation of the branching ratio BR($H_5^0 \to \gamma\gamma$) for $m_5 < 2M_W$, where the $Z\gamma$ decay can contribute non-negligibly to the $H_5^0$ total width.  We now use this to apply a new constraint on the GM model from a LEP search for $e^+e^- \to Z H_5^0$ with $H_5^0 \to \gamma\gamma$~\cite{LEP2002}.  We take the numerical exclusion limit from HiggsBounds 4.2.0~\cite{HiggsBounds}.  The exclusion is shown in the left panel of Fig.~\ref{fig:LEP} as a limit on $(\kappa_Z^{H_5^0})^2 \times {\rm BR}(H_5^0 \to \gamma\gamma)$ as a function of $m_5$, where $\kappa_Z^{H_5^0} = 2 \sin\theta_H/\sqrt{3}$ is the $H_5^0ZZ$ coupling normalized to that of the SM Higgs boson.  The points above the blue curve are excluded, and will be colored red in all the plots in this section.  The black points are allowed by all constraints considered in this section.

The effect of the LEP $H_5^0 \to \gamma\gamma$ constraint on the GM model parameter space can be better understood by studying the $m_5$--$\sin\theta_H$ plane, as shown in the right panel of Fig.~\ref{fig:LEP}.  To illustrate the effects of the other experimental constraints on the model, we show the points excluded by $b \to s \gamma$ in cyan and the points excluded by the ATLAS like-sign $WW$ cross section in green.  Again the red points are excluded by the LEP $H_5^0 \to \gamma\gamma$ constraint and the black points are allowed.  We see that LEP excludes most of the parameter space for $m_5 \lesssim 110$~GeV, except for points at low $\sin\theta_H$ for which the $e^+e^- \to Z H_5^0$ cross section is suppressed and a smattering of points at higher $\sin\theta_H$ for which BR($H_5^0 \to \gamma\gamma$) is suppressed due to cancellations among the loop amplitudes.

\begin{figure}
\begin{center}
\resizebox{0.49\textwidth}{!}{\includegraphics{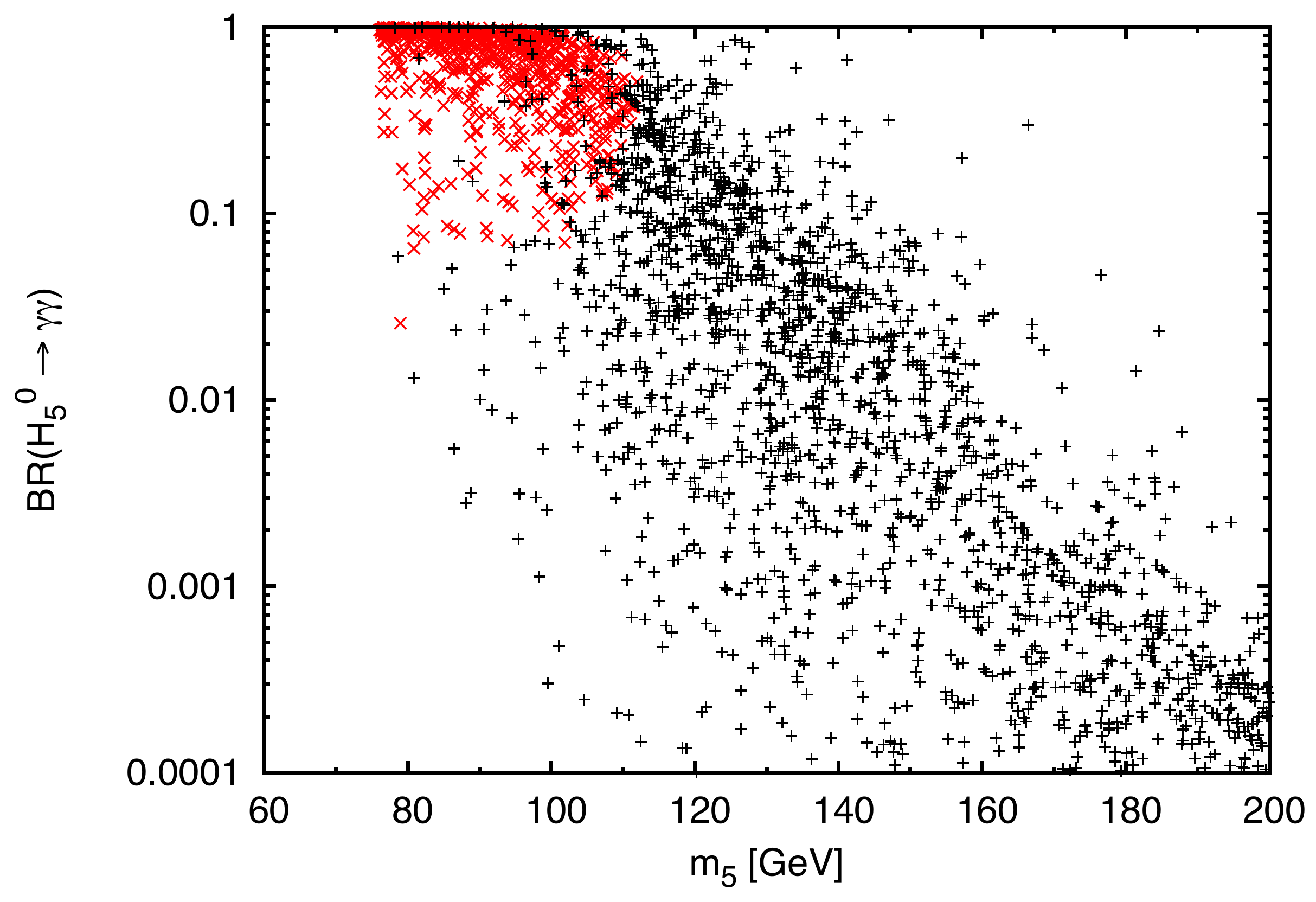}}
\resizebox{0.49\textwidth}{!}{\includegraphics{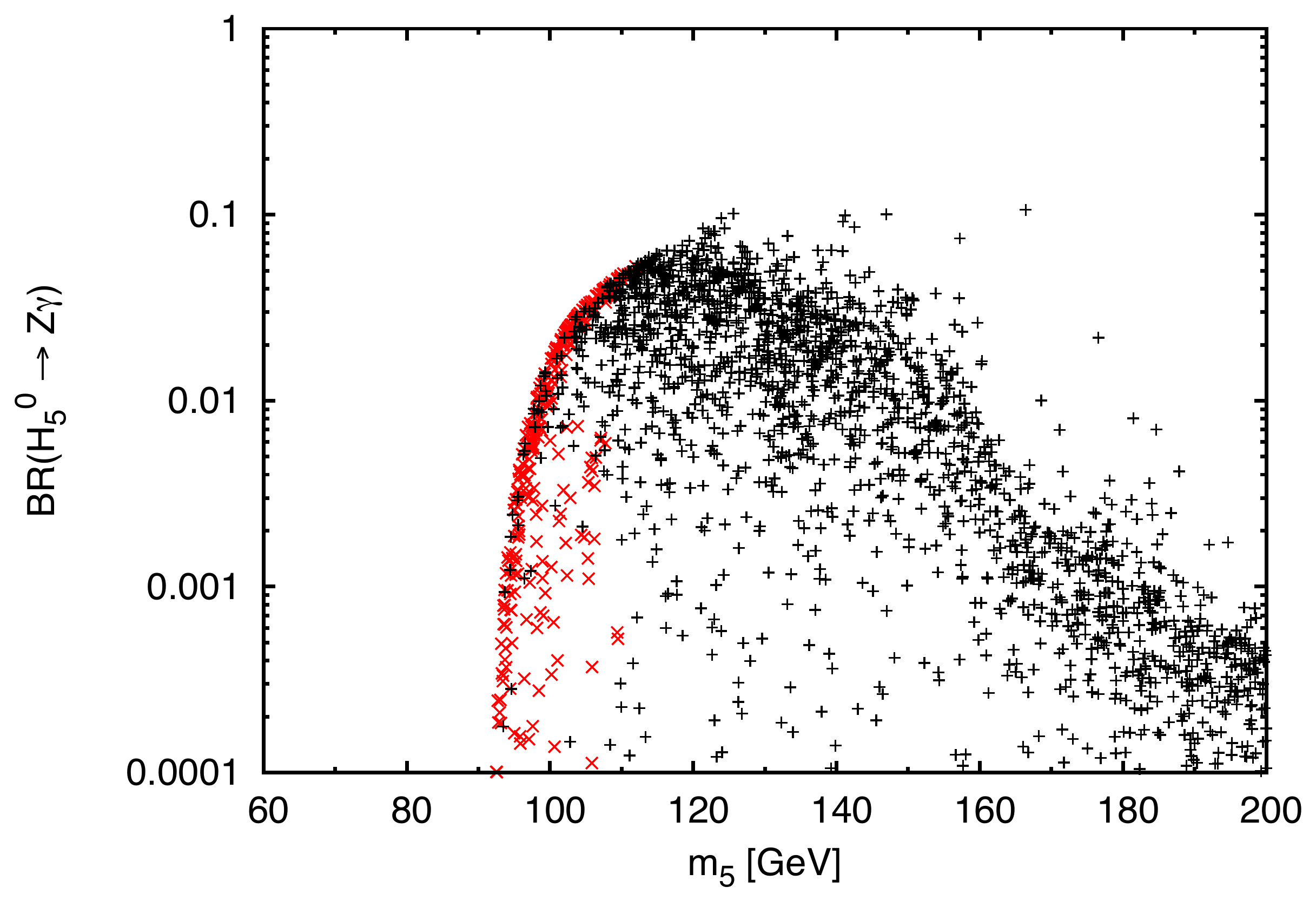}}
\caption{Branching ratios of $H_5^0$ to $\gamma\gamma$ (left) and $Z \gamma$ (right) as a function of its mass $m_5$.  Red points are excluded by the LEP search for $e^+e^- \to ZH$, $H \to \gamma\gamma$.  Black points are allowed by all constraints.}
\label{fig:BRsHvsm5}
\end{center}
\end{figure}

In Fig.~\ref{fig:BRsHvsm5} we show the branching ratios of $H_5^0 \to \gamma\gamma$ (left panel) and $H_5^0 \to Z\gamma$ (right panel) as a function of $m_5$.  The black points are allowed.  We see that BR($H_5^0 \to \gamma\gamma$) can reach several tens of percent for $m_5 \lesssim 130$~GeV, and be above 1\% for a large fraction of the parameter space with $m_5 \lesssim 2M_W$.  Similarly, BR($H_5^0 \to Z\gamma$) can reach several percent for $m_5 \sim 110$--150~GeV, but never surpasses 10\%.  The rapid decline of BR($H_5^0 \to Z \gamma$) for $m_5 \lesssim 110$~GeV is due to the kinematic suppression from the on-shell $Z$ boson.

\begin{figure}
\begin{center}
\resizebox{0.75\textwidth}{!}{\includegraphics{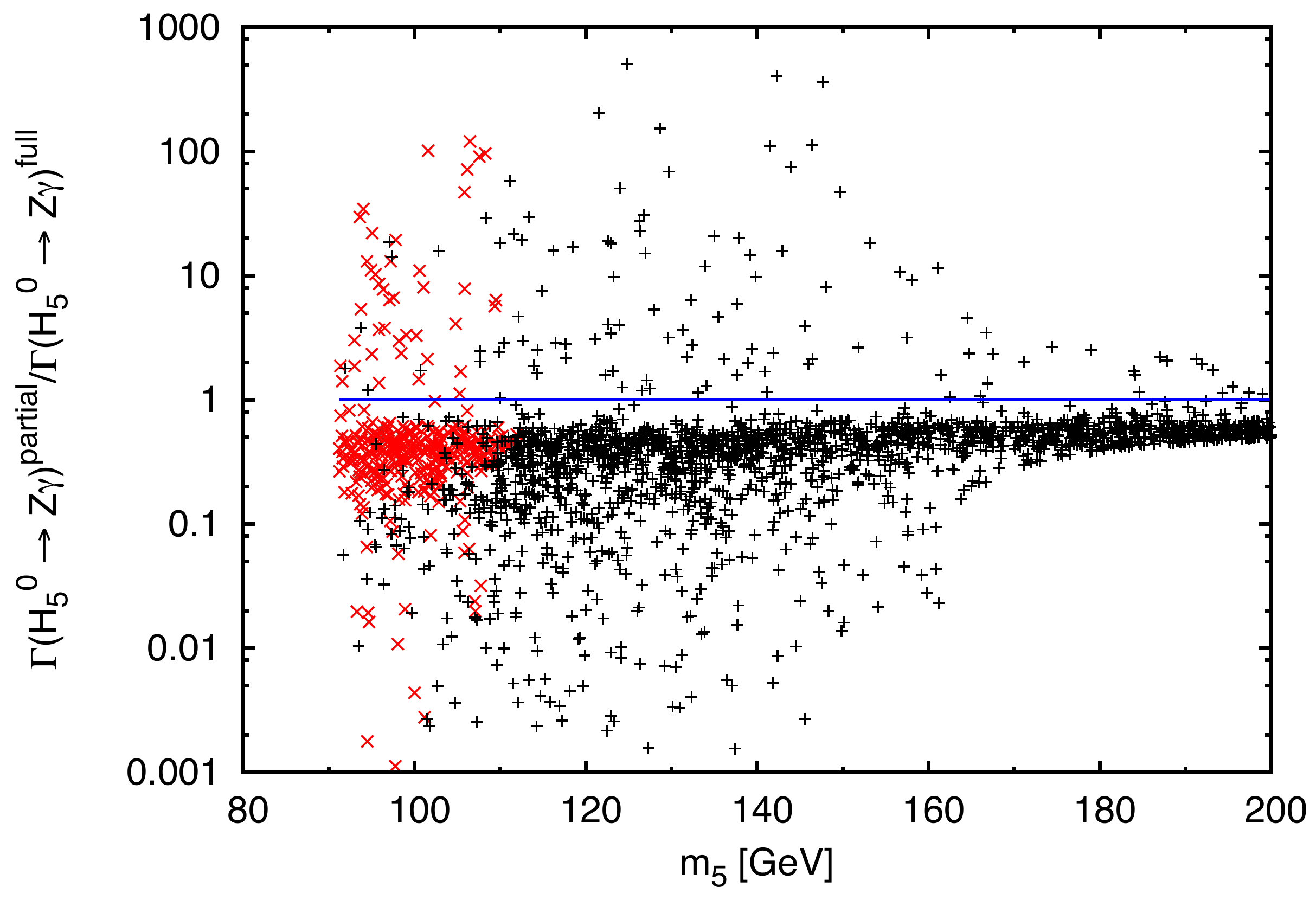}}
\caption{Comparison of a ``partial'' calculation of $\Gamma(H_5^0 \to Z \gamma)$ including only the SM-like diagrams to the full calculation.  Red points are excluded by the LEP search for $e^+e^- \to ZH$, $H \to \gamma\gamma$.  Black points are allowed by all constraints.}
\label{fig:H50Zga-ratio}
\end{center}
\end{figure}

In Fig.~\ref{fig:H50Zga-ratio} we study the effect of the new vector-scalar-scalar and scalar-vector-vector contributions to $H_5^0 \to Z \gamma$. In this figure we plot a ``partial'' calculation of $\Gamma(H_5^0 \to Z\gamma)$, obtained by computing only the usual $W$ and scalar loop diagrams for which standard expressions are available~\cite{HHG}, normalized to the full calculation of Eq.~(\ref{eq:Ahgz}).  Over most of the parameter space, neglecting the new vector-scalar-scalar and scalar-vector-vector loop contributions would lead to a result for $\Gamma(H_5^0 \to Z\gamma)$ about a factor of two smaller than that of the full calculation, except at parameter points where an accidental cancellation among loop amplitudes occurs in either the ``partial'' or the full result.

\begin{figure}
\begin{center}
\resizebox{0.49\textwidth}{!}{\includegraphics{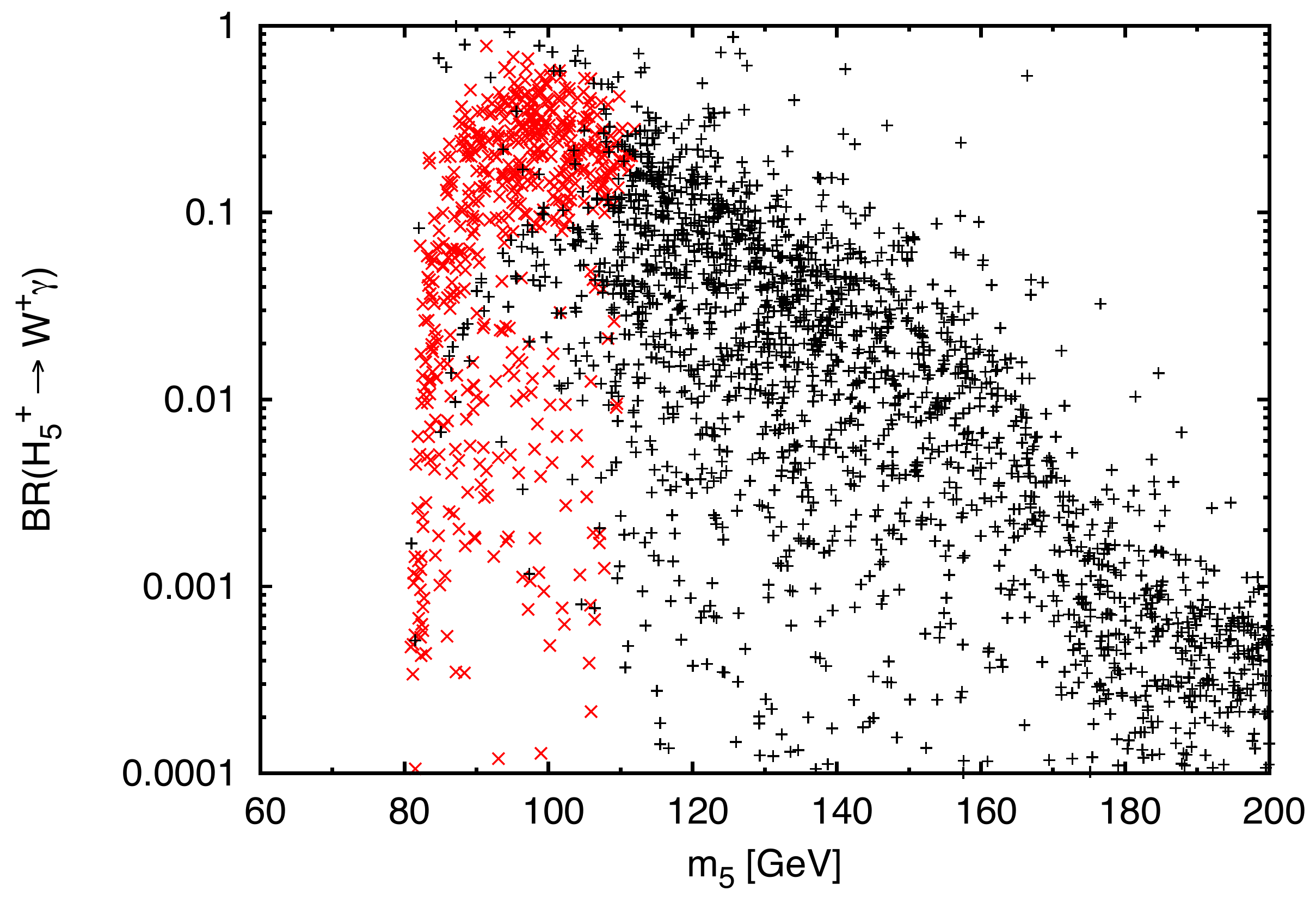}}
\resizebox{0.49\textwidth}{!}{\includegraphics{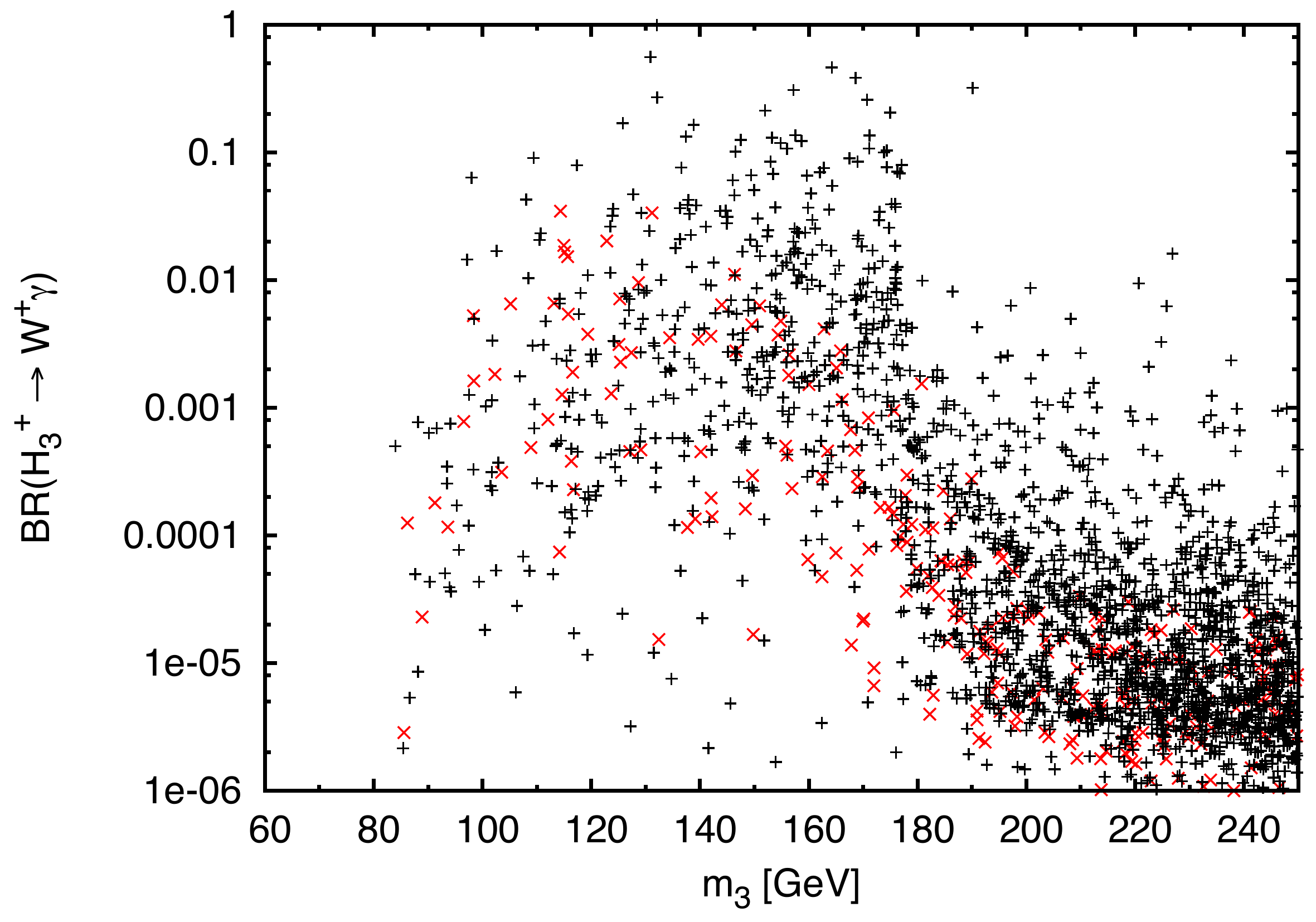}}
\caption{Branching ratios of $H_5^+ \to W^+ \gamma$ as a function of $m_5$ (left) and of $H_3^+ \to W^+ \gamma$ as a function of $m_3$ (right).  Red points are excluded by the LEP search for $e^+e^- \to ZH$, $H \to \gamma\gamma$.  Black points are allowed by all constraints.}
\label{fig:WgamBRvsmS}
\end{center}
\end{figure}

In Fig.~\ref{fig:WgamBRvsmS} we plot the branching ratios for $H_5^+ \to W^+\gamma$ (left) and $H_3^+ \to W^+ \gamma$ (right) as a function of $m_5$ and $m_3$, respectively.  The black points are allowed.  We see that BR($H_5^+ \to W^+ \gamma$) can reach a few tens of percent for $m_5 \lesssim 130$~GeV, and be above 1\% for a large fraction of the parameter space with $m_5 \lesssim 2M_W$.  BR($H_3^+ \to W^+ \gamma$) is typically smaller due to competition with decays to fermion pairs, though it can reach tens of percent for select parameter points.  This happens because $H_3^+ \to W^+ \gamma$ receives contributions from scalar loop diagrams (see Table~\ref{tab:H3ppart}), which can remain unsuppressed at small $\sin\theta_H$ when the tree-level decays of $H_3^+$ into fermion pairs are suppressed.

Finally we comment on the decays of $H_3^0$.  At low masses, the decays of this state are dominated by $f \bar f$ and $gg$, as well as decays to $Zh$, $ZH$, $ZH_5^0$, and/or $W^{\pm}H_5^{\mp}$ when not too kinematically suppressed.  Because $H_3^0$ is CP-odd, its loop-induced decays to $\gamma\gamma$ and $Z\gamma$ receive contributions only from fermions in the loop.  Therefore the partial widths for these loop-induced decays, as well as the loop-induced decay to $gg$ and the tree-level decay to $f \bar f$, all scale with the same $H_3^0 \bar f f$ coupling modification factor $\tan^2 \theta_H$.  None of these decays of $H_3^0$ involve the new one-loop diagrams computed in this paper, and they are already implemented in GMCALC~1.2.0.

\section{Conclusions}
\label{sec:conclusions}

In this paper we evaluated the one-loop contributions to $H_i \to V \gamma$ from ``heterogeneous'' loop diagrams involving particles with two different masses propagating in the loop.  These are necessary for a full leading-order calculation of the decay widths of $H_3^+ \to W^+ \gamma$, $H_5^+ \to W^+ \gamma$, and $H_5^0 \to Z \gamma$ in the GM model.  The novel results presented here are for (1) the scalar loop diagram with $m_{H_i} \neq m_{s_1} \neq m_{s_2}$, which contributes to $H_3^+ \to W^+ \gamma$; (2) the vector-scalar-scalar loop diagram, which contributes to $H_5^0 \to Z \gamma$ and $H_5^+ \to W^+ \gamma$; (3) the scalar-vector-vector loop diagram with $M_V \neq M_{X_2}$, which contributes to $H_5^0 \to Z \gamma$; and (4) the vector loop diagram which contributes to $H_5^+ \to W^+ \gamma$.  We gave the results for these diagrams in terms of the LoopTools functions for ease of numerical implementation.  We also recalculated the heterogeneous loop diagrams previously computed in Ref.~\cite{Ilisie:2014hea} in order to give expressions for them in terms of the LoopTools functions.

Using these results we performed numerical scans over the theoretically- and experimentally-allowed parameter space of the GM model in order to study the behavior of the $H_i \to V \gamma$ branching ratios.  We showed that a LEP search for $e^+e^- \to Z H_5^0$ with $H_5^0 \to \gamma\gamma$ strongly constrains the GM model parameter space for $m_5 \lesssim 110$~GeV.  

Our results for the loop-induced $H_i \to V \gamma$ decays will be implemented into GMCALC~1.3.0 and higher, which will allow the experimental searches for $H_5^0$ and $H_5^+$ at the LHC to be reliably extended below the $VV$ threshold.

\begin{acknowledgments}
C.D.\ was a Durham International Junior Research Fellow and has been supported in part by the Research Executive Agency of the European Union under Grant No.\ PITN-GA-2012-315877 (MC-Net). 
K.H.\ and H.E.L.\ were supported by the Natural Sciences and Engineering Research Council of Canada.  K.H.\ was also supported by the Government of Ontario through an Ontario Graduate Scholarship.  H.E.L.\ acknowledges additional support from the grant H2020-MSCA-RISE-2014 No.\ 645722 (NonMinimalHiggs) and thanks the CERN Theory Division for hospitality during part of this work.
\end{acknowledgments}

\appendix

\section{The Georgi-Machacek model}
\label{sec:GMmodel}

The scalar sector of the GM model~\cite{Georgi:1985nv,Chanowitz:1985ug} consists of the usual complex doublet $(\phi^+,\phi^0)^T$ with hypercharge\footnote{We use $Q = T^3 + Y/2$.} $Y = 1$, a real triplet $(\xi^+,\xi^0, -\xi^{+*})^T$ with $Y = 0$, and  a complex triplet $(\chi^{++},\chi^+,\chi^0)^T$ with $Y=2$.  The doublet is responsible for the fermion masses as in the SM.
Custodial symmetry is preserved at tree level by imposing a global SU(2)$_L \times$SU(2)$_R$ symmetry on the scalar potential.
In order to make this symmetry explicit, we write the doublet in the form of a bidoublet $\Phi$ and combine the triplets to form a bitriplet $X$:
\begin{equation}
	\Phi = \left( \begin{array}{cc}
	\phi^{0*} &\phi^+  \\
	-\phi^{+*} & \phi^0  \end{array} \right), \qquad
	X =
	\left(
	\begin{array}{ccc}
	\chi^{0*} & \xi^+ & \chi^{++} \\
	 -\chi^{+*} & \xi^{0} & \chi^+ \\
	 \chi^{++*} & -\xi^{+*} & \chi^0  
	\end{array}
	\right).
	\label{eq:PX}
\end{equation}
The vevs are defined by $\langle \Phi  \rangle = \frac{ v_{\phi}}{\sqrt{2}} I_{2\times2}$  and $\langle X \rangle = v_{\chi} I_{3 \times 3}$, where $I$ is the unit matrix and the $W$ and $Z$ boson masses constrain
\begin{equation}
	v_{\phi}^2 + 8 v_{\chi}^2 \equiv v^2 = \frac{1}{\sqrt{2} G_F} \approx (246~{\rm GeV})^2.
	\label{eq:vevrelation}
\end{equation} 

The most general gauge-invariant scalar potential involving these fields that conserves custodial SU(2) is given, in the conventions of Ref.~\cite{Hartling:2014zca}, by\footnote{A translation table to other parameterizations in the literature has been given in the appendix of Ref.~\cite{Hartling:2014zca}.}
\begin{eqnarray}
	V(\Phi,X) &= & \frac{\mu_2^2}{2}  \text{Tr}(\Phi^\dagger \Phi) 
	+  \frac{\mu_3^2}{2}  \text{Tr}(X^\dagger X)  
	+ \lambda_1 [\text{Tr}(\Phi^\dagger \Phi)]^2  
	+ \lambda_2 \text{Tr}(\Phi^\dagger \Phi) \text{Tr}(X^\dagger X)   \nonumber \\
          & & + \lambda_3 \text{Tr}(X^\dagger X X^\dagger X)  
          + \lambda_4 [\text{Tr}(X^\dagger X)]^2 
           - \lambda_5 \text{Tr}( \Phi^\dagger \tau^a \Phi \tau^b) \text{Tr}( X^\dagger t^a X t^b) 
           \nonumber \\
           & & - M_1 \text{Tr}(\Phi^\dagger \tau^a \Phi \tau^b)(U X U^\dagger)_{ab}  
           -  M_2 \text{Tr}(X^\dagger t^a X t^b)(U X U^\dagger)_{ab}.
           \label{eq:potential}
\end{eqnarray} 
Here the SU(2) generators for the doublet representation are $\tau^a = \sigma^a/2$ with $\sigma^a$ being the Pauli matrices, the generators for the triplet representation are
\begin{equation}
	t^1= \frac{1}{\sqrt{2}} \left( \begin{array}{ccc}
	 0 & 1  & 0  \\
	  1 & 0  & 1  \\
	  0 & 1  & 0 \end{array} \right), \qquad  
	  t^2= \frac{1}{\sqrt{2}} \left( \begin{array}{ccc}
	 0 & -i  & 0  \\
	  i & 0  & -i  \\
	  0 & i  & 0 \end{array} \right), \qquad 
	t^3= \left( \begin{array}{ccc}
	 1 & 0  & 0  \\
	  0 & 0  & 0  \\
	  0 & 0 & -1 \end{array} \right),
\end{equation}
and the matrix $U$, which rotates $X$ into the Cartesian basis, is given by~\cite{Aoki:2007ah}
\begin{equation}
	 U = \left( \begin{array}{ccc}
	- \frac{1}{\sqrt{2}} & 0 &  \frac{1}{\sqrt{2}} \\
	 - \frac{i}{\sqrt{2}} & 0  &   - \frac{i}{\sqrt{2}} \\
	   0 & 1 & 0 \end{array} \right).
	 \label{eq:U}
\end{equation}

The physical fields can be organized by their transformation properties under the custodial SU(2) symmetry into a fiveplet, a triplet, and two singlets.  The fiveplet and triplet states are given by
\begin{eqnarray}
	&&H_5^{++} = \chi^{++}, \qquad
	H_5^+ = \frac{\left(\chi^+ - \xi^+\right)}{\sqrt{2}}, \qquad
	H_5^0 = \sqrt{\frac{2}{3}} \xi^0 - \sqrt{\frac{1}{3}} \chi^{0,r}, \nonumber \\
	&&H_3^+ = - s_H \phi^+ + c_H \frac{\left(\chi^++\xi^+\right)}{\sqrt{2}}, \qquad
	H_3^0 = - s_H \phi^{0,i} + c_H \chi^{0,i},
\end{eqnarray}
where the vevs are parameterized by
\begin{equation}
	c_H \equiv \cos\theta_H = \frac{v_{\phi}}{v}, \qquad
	s_H \equiv \sin\theta_H = \frac{2\sqrt{2}\,v_\chi}{v},
\end{equation}
and we have decomposed the neutral fields into real and imaginary parts according to
\begin{equation}
	\phi^0 \to \frac{v_{\phi}}{\sqrt{2}} + \frac{\phi^{0,r} + i \phi^{0,i}}{\sqrt{2}},
	\qquad
	\chi^0 \to v_{\chi} + \frac{\chi^{0,r} + i \chi^{0,i}}{\sqrt{2}}, 
	\qquad
	\xi^0 \to v_{\chi} + \xi^0.
\end{equation}
The masses within each custodial multiplet are degenerate at tree level and can be written (after eliminating $\mu_2^2$ and $\mu_3^2$ in favor of the vevs) as\footnote{Note that the ratio $M_1/v_{\chi}$ can be written using the minimization condition $\partial V/ \partial v_{\chi} = 0$ as
\begin{equation}
	\frac{M_1}{v_{\chi}} = \frac{4}{v_{\phi}^2} 
	\left[ \mu_3^2 + (2 \lambda_2 - \lambda_5) v_{\phi}^2 
	+ 4(\lambda_3 + 3 \lambda_4) v_{\chi}^2 - 6 M_2 v_{\chi} \right],
\end{equation}
which is finite in the limit $v_{\chi} \to 0$.}
\begin{eqnarray}
	m_5^2 &=& \frac{M_1}{4 v_{\chi}} v_\phi^2 + 12 M_2 v_{\chi} 
	+ \frac{3}{2} \lambda_5 v_{\phi}^2 + 8 \lambda_3 v_{\chi}^2, \nonumber \\
	m_3^2 &=&  \frac{M_1}{4 v_{\chi}} (v_\phi^2 + 8 v_{\chi}^2) 
	+ \frac{\lambda_5}{2} (v_{\phi}^2 + 8 v_{\chi}^2) 
	= \left(  \frac{M_1}{4 v_{\chi}} + \frac{\lambda_5}{2} \right) v^2.
\end{eqnarray}
Note that the custodial-fiveplet states $H_5$ consist entirely of the triplet fields, and hence do not couple to fermions at tree level.  In contrast, the $H_3$ states contain a doublet admixture and hence do couple to fermions.

The two custodial-singlet mass eigenstates are given by
\begin{equation}
	h = c_{\alpha} \, \phi^{0,r} - s_{\alpha} \, H_1^{0\prime},  \qquad
	H = s_{\alpha} \, \phi^{0,r} + c_{\alpha} \, H_1^{0\prime},
	\label{mh-mH}
\end{equation}
where $c_{\alpha} = \cos\alpha$, $s_{\alpha} = \sin\alpha$, and
\begin{equation}
	H_1^{0 \prime} = \sqrt{\frac{1}{3}} \xi^0 + \sqrt{\frac{2}{3}} \chi^{0,r}.
\end{equation}
The mixing angle and masses are given by
\begin{eqnarray}
	&&\sin 2 \alpha =  \frac{2 \mathcal{M}^2_{12}}{m_H^2 - m_h^2},    \qquad
	\cos 2 \alpha =  \frac{ \mathcal{M}^2_{22} - \mathcal{M}^2_{11}  }{m_H^2 - m_h^2}, 
	\nonumber \\
	&&m^2_{h,H} = \frac{1}{2} \left[ \mathcal{M}_{11}^2 + \mathcal{M}_{22}^2
	\mp \sqrt{\left( \mathcal{M}_{11}^2 - \mathcal{M}_{22}^2 \right)^2 
	+ 4 \left( \mathcal{M}_{12}^2 \right)^2} \right],
	\label{eq:hmass}
\end{eqnarray}
where we choose $m_h < m_H$, and 
\begin{eqnarray}
	\mathcal{M}_{11}^2 &=& 8 \lambda_1 v_{\phi}^2, \nonumber \\
	\mathcal{M}_{12}^2 &=& \frac{\sqrt{3}}{2} v_{\phi} 
	\left[ - M_1 + 4 \left(2 \lambda_2 - \lambda_5 \right) v_{\chi} \right], \nonumber \\
	\mathcal{M}_{22}^2 &=& \frac{M_1 v_{\phi}^2}{4 v_{\chi}} - 6 M_2 v_{\chi} 
	+ 8 \left( \lambda_3 + 3 \lambda_4 \right) v_{\chi}^2.
\end{eqnarray}

The couplings of the scalars in the GM model that we use in this paper are collected in Appendix~\ref{app:couplings}.

\section{Feynman rules for the GM model}
\label{app:couplings}

Here we summarize the Feynman rules for the GM model that we have used in the one-loop decay calculations.  A full set of Feynman rules can be found in Ref.~\cite{Hartling:2014zca}.  In what follows, all particles and momenta are incoming.  For the covariant derivative we use the sign convention $\mathcal{D}_{\mu} = \partial_{\mu} - i g A_{\mu}^a T^a$.

\subsection{Scalar couplings to fermions \label{A:GMFRfermion}}

The Feynman rules for the vertices involving a neutral scalar and two fermions are given as follows:
\begin{eqnarray}
	h \bar f f: &\quad& -i \frac{m_f}{v} \frac{\cos \alpha}{\cos \theta_H} = i\,C_{h \bar f f}, \nonumber \\
	H \bar f f: &\quad& -i \frac{m_f}{v} \frac{\sin \alpha}{\cos \theta_H} = i\,C_{H \bar f f}, \nonumber \\
	H_3^0 \bar u u: &\quad& \frac{m_u}{v} \tan \theta_H \gamma_5, = C_{H_3^0 \bar u u} \gamma_5\nonumber \\
	H_3^0 \bar d d: &\quad& -\frac{m_d}{v} \tan \theta_H \gamma_5 = C_{H_3^0 \bar d d}\gamma_5.
\end{eqnarray}
Here $f$ stands for any charged fermion, $u$ stands for any up-type quark, and $d$ stands for any down-type quark or charged lepton.  

The Feynman rules for the vertices involving a charged scalar and two fermions are given as follows, with all particles incoming:
\begin{eqnarray}
	H_3^+ \bar u d: &\quad& -i \frac{\sqrt{2}}{v} V_{ud} \tan\theta_H
		\left( m_u P_L - m_d P_R \right), \nonumber \\
	H_3^{+*} \bar d u: &\quad& -i \frac{\sqrt{2}}{v} V_{ud}^* \tan\theta_H 
		\left( m_u P_R - m_d P_L \right).
\end{eqnarray}
Here $V_{ud}$ is the appropriate element of the CKM matrix and the projection operators are defined as $P_{R,L} = (1 \pm \gamma_5)/2$. We define the coupling coefficients $C^S$ and $C^P$ used in the work above according to $i\left(C_{H_3^+\bar f_1 f_2}^S + C_{H_3^+\bar f_1 f_2}^P\gamma_5\right)$. 

The custodial fiveplet states do not couple to fermions.

\subsection{Triple scalar couplings}

The Feynman rules for the triple-scalar couplings involving incoming scalars $s_1 s_2 s_3$ are given by $-i C_{s_1s_2 s_3}$, with all particles incoming.  The ordering of the indices does not matter for these couplings.  The couplings used in our calculations given as follows:
\begin{eqnarray}
	C_{H_3^+H_3^{+*}h} &=& \frac{1}{\sqrt{3} v^2} 
	\left\{ \sqrt{3} c_{\alpha} \left[ (4 \lambda_2 - \lambda_5) v_{\phi}^3 + 8 (8 \lambda_1 + \lambda_5) v_{\phi} v_{\chi}^2 + 4 M_1 v_{\phi} v_{\chi} \right] \right. \nonumber \\
	&& \left. - s_{\alpha} \left[ 8 (\lambda_3 + 3 \lambda_4 + \lambda_5) v_{\phi}^2 v_{\chi}
	+ 16 (6 \lambda_2 + \lambda_5) v_{\chi}^3 + 4 M_1 v_{\chi}^2 - 6 M_2 v_{\phi}^2 \right] \right\}, \\
	C_{H_3^+H_3^{+*}H} &=& \frac{1}{\sqrt{3} v^2}
	\left\{ \sqrt{3} s_{\alpha} \left[ (4 \lambda_2 - \lambda_5) v_{\phi}^3 + 8 (8 \lambda_1 + \lambda_5) v_{\phi} v_{\chi}^2 + 4 M_1 v_{\phi} v_{\chi} \right] \right. \nonumber \\
	&& \left. + c_{\alpha} \left[ 8 (\lambda_3 + 3 \lambda_4 + \lambda_5) v_{\phi}^2 v_{\chi}
	+ 16 (6 \lambda_2 + \lambda_5) v_{\chi}^3 + 4 M_1 v_{\chi}^2 - 6 M_2 v_{\phi}^2 \right] \right\}, \\
	C_{H_5^+H_5^{+*}h} = C_{H_5^{++}H_5^{++*}h} &=& 
	c_{\alpha} \left[ (4 \lambda_2 + \lambda_5) v_{\phi} \right]
	- \sqrt{3} s_{\alpha} \left[ 8 (\lambda_3 + \lambda_4) v_{\chi} + 2 M_2 \right], \\
	C_{H_5^+H_5^{+*}H} = C_{H_5^{++}H_5^{++*}H} &=&
	s_{\alpha} \left[ (4 \lambda_2 + \lambda_5) v_{\phi} \right]
	+ \sqrt{3} c_{\alpha} \left[ 8 (\lambda_3 + \lambda_4) v_{\chi} + 2 M_2 \right], \\
	C_{H_3^+H_3^{+*}H_5^0} &=& \sqrt{\frac{2}{3}} \frac{1}{v^2}
	\left[ 2(\lambda_3 - 2 \lambda_5) v_\phi^2 v_\chi - 8 \lambda_5 v_\chi^3
	+ 4 M_1 v_\chi^2 + 3 M_2 v_\phi^2 \right],\\
	C_{H_3^0H_3^+H_5^{+*}} &=& -i\frac{\sqrt{2}}{v^2}
	\left[ 2(\lambda_3 - 2 \lambda_5) v_\phi^2 v_\chi - 8 \lambda_5 v_\chi^3
	+ 4 M_1 v_\chi^2 + 3 M_2 v_\phi^2 \right],  \\
	C_{H_3^+H_3^+H_5^{++*}} &=& -\frac{2}{v^2} 
	\left[ 2(\lambda_3 - 2 \lambda_5) v_\phi^2 v_\chi - 8 \lambda_5 v_\chi^3
	+ 4 M_1 v_\chi^2 + 3 M_2 v_\phi^2 \right],  \\
	C_{H_5^+H_5^{+*}H_5^0} &=& \sqrt{6} \left( 2 \lambda_3 v_\chi - M_2 \right),  \\
	C_{H_5^+H_5^{+}H_5^{++*}} &=& -6 \left( 2 \lambda_3 v_\chi -  M_2 \right),  \\
	C_{H_5^{++}H_5^{++*}H_5^0} &=& -2 \sqrt{6} \left( 2 \lambda_3 v_\chi - M_2 \right). 
\end{eqnarray}

\subsection{Scalar-vector-vector couplings}

The Feynman rules for the vertices involving a scalar and two gauge bosons are defined as $i e^2 C_{sV_1V_2} g^{\mu \nu}$. The couplings used in our calculations are given by
\begin{eqnarray}
	C_{hW^+W^{+*}}  &=& c_W^2 C_{hZZ} 
	= -\frac{1}{6 s_W^2} \left( 8\sqrt{3} s_\alpha v_\chi - 3 c_\alpha v_\phi\right),   \\
	C_{HW^+W^{+*}}  &=&   c_W^2 C_{HZZ} 
	=   \frac{1}{6 s_W^2} \left( 8\sqrt{3} c_\alpha v_\chi + 3 s_\alpha v_\phi\right), \\
	C_{H_5^0 W^+ W^{+*}} &=& \sqrt{\frac{2}{3}} \frac{1}{s_W^2} v_{\chi},   \\
	C_{H_5^+ W^{+*} Z} &=&   -\frac{\sqrt{2}}{c_W s_W^2} v_{\chi},  \\
	C_{H_5^{++} W^{+*} W^{+*}} &=&   \frac{2}{s_W^2} v_{\chi}. 
\end{eqnarray}

\subsection{Vector-scalar-scalar couplings}

The Feynman rules for the vertices involving two scalars and a single $Z$ boson are defined as
$i e C_{Z s_1s_2} \left(p_1 - p_2\right)_\mu$,
where $p_1$ ($p_2$) is the incoming momentum of incoming scalar $s_1$ ($s_2$).  The couplings are given by
\begin{eqnarray}
	C_{ZhH_3^0} &=& -i\sqrt{\frac{2}{3}}\frac{1}{s_Wc_W}\left(\sqrt{3}\frac{v_\chi}{v} c_\alpha + s_\alpha \frac{v_\phi}{v} \right),  \\
	C_{ZHH_3^0} &=& i\sqrt{\frac{2}{3}}\frac{1}{s_Wc_W}\left(c_\alpha \frac{v_\phi}{v} - \sqrt{3}\frac{v_\chi}{v} s_\alpha \right),  \\
	C_{ZH_3^0 H_5^0} &=& i \sqrt{\frac{1}{3}} \frac{1}{s_Wc_W}\frac{v_\phi}{v},  \\
	C_{ZH_3^+H_3^{+*}} &=& C_{ZH_5^+H_5^{+*}} 
	= \frac{1}{2 s_Wc_W}\left(1 - 2s_W^2\right),  \\
	C_{ZH_5^{++}H_5^{++*}} &=& \frac{1}{s_Wc_W}\left(1 - 2 s_W^2\right),  \\
	C_{ZH_3^+H_5^{+*}} &=& -\frac{1}{2 s_Wc_W}\frac{v_\phi}{v}.
\end{eqnarray}

The Feynman rules for the vertices involving two scalars and a single $W^+$ boson are defined as
$i e C_{W^+ s_1s_2} \left(p_1 - p_2\right)_\mu$,
where again $p_1$ ($p_2$) is the incoming momentum of incoming scalar $s_1$ ($s_2$). The couplings are given by
\begin{eqnarray}
	C_{W^+ h H_3^{+*}}  &=&     - \sqrt{\frac{2}{3}} \frac{1}{  s_W} 
	\left( \sqrt{3} c_\alpha \frac{v_\chi}{v} +  s_\alpha  \frac{v_\phi}{v}    \right),       
\\
	C_{W^+ H H_3^{+*}}  &=&     - \sqrt{\frac{2}{3}} \frac{1}{  s_W} 
	\left( \sqrt{3} s_\alpha \frac{v_\chi}{v} -  c_\alpha  \frac{v_\phi}{v}    \right),        \\
	C_{W^+  H_3^0 H_3^{+*}}   &=&  - \frac{i}{2} \frac{1}{s_W}	,	 \\
	C_{W^+ H_3^{+*} H_5^0 }   &=&   - \frac{1}{2 \sqrt{3}} \frac{1}{s_W} \frac{v_\phi}{v},		
	 \\
	C_{W^+ H_3^0 H_5^{+*}}   &=&     \frac{i}{2} \frac{1}{s_W} \frac{v_\phi}{v},		\\
	C_{W^+ H_3^+ H_5^{++*} }   &=&      \frac{1}{\sqrt{2}} \frac{1}{s_W} \frac{v_\phi}{v},
	 \\
	C_{W^+ H_5^{+*} H_5^0}   &=&     \frac{\sqrt{3}}{2} \frac{1}{s_W}, 	 \\
	C_{W^+ H_5^+ H_5^{++*}}   &=&  	 \frac{1}{\sqrt{2}} \frac{1}{s_W}.	  
\end{eqnarray}

The couplings for the conjugate processes involving an incoming $W^-$ are obtained using
\begin{equation}
	C_{W^- s_1^* s_2^*} = - C_{W^+ s_1 s_2}^*.
\end{equation}

\subsection{Couplings involving Goldstone bosons}
\label{app:goldstonecoups}

Our calculation of the vector-scalar-scalar, scalar-vector-vector, and vector loop diagrams in the 't~Hooft-Feynman gauge require the calculation of diagrams involving Goldstone bosons.  We collect the relevant couplings here.

The couplings of Goldstone bosons to other scalars are given by $-i C_{s_1 s_2 s_3}$, where the coefficients used in this paper are
\begin{align}
C_{G^0 H_5^+ H_3^{+*}} &= i \frac{v_\phi}{v^2}\left(m_5^2-m_3^2\right) \,, \\
C_{G^+ H_3^{+*} H_5^{0}} &= \frac{1}{\sqrt{3}}\frac{v_\phi}{v^2}\left(m_5^2-m_3^2\right) \,,\\
C_{G^+ H_5^{+*} H_3^0} &= -i \frac{v_\phi}{v^2}\left(m_5^2-m_3^2\right) \,,\\
C_{G^+ H_3^{+} H_5^{++*}} &= -\sqrt{2} \frac{v_\phi}{v^2}\left(m_5^2-m_3^2\right) \,,\\
C_{G^0 H_5^+ G^{+*}} &= \frac{i }{\sqrt{2} v^2} \left(32\lambda_3 v_\chi^3 + 6 \lambda_5 v_\phi^2 v_\chi + M_1 v_\phi^2 + 48 M_2 v_\chi^2\right) \,,\\
C_{G^+ G^{+*} H_5^0} &= \frac{1}{\sqrt{6} v^2} \left(32\lambda_3 v_\chi^3 + 6 \lambda_5 v_\phi^2 v_\chi + M_1 v_\phi^2 + 48 M_2 v_\chi^2\right) \,,\\
C_{G^+ G^{+} H_5^{++*}} &= - \frac{1}{v^2} \left(32\lambda_3 v_\chi^3 + 6 \lambda_5 v_\phi^2 v_\chi + M_1 v_\phi^2 + 48 M_2 v_\chi^2\right) \,, \\
C_{G^+ H_3^{+*} h} & = -\frac{c_\alpha}{\sqrt{2} v^2}\left[2 v_\phi^2 v_\chi (16\lambda_1 +3\lambda_5 - 8\lambda_2) - 16\lambda_5 v_\chi^3+M_1 (v_\phi^2-8 v_\chi^2)\right] \nonumber \\
                       & \quad + \sqrt{\frac{2}{3}}\frac{v_\phi s_\alpha}{v^2}\left[4 v_\chi^2(6 \lambda_2-4 \lambda_3 - 12 \lambda_4-\lambda_5 )+v_\chi (M_1+12 M_2) +\lambda_5 v_\phi^2\right]\,, \\
C_{G^+ H_3^{+*} H} & =  -\frac{s_\alpha}{\sqrt{2} v^2}\left[2 v_\phi^2 v_\chi (16\lambda_1 +3\lambda_5 - 8\lambda_2) - 16\lambda_5 v_\chi^3+M_1 (v_\phi^2-8 v_\chi^2)\right] \nonumber \\
                       & \quad - \sqrt{\frac{2}{3}}\frac{v_\phi c_\alpha}{v^2}\left[4 v_\chi^2(6 \lambda_2-4 \lambda_3 - 12 \lambda_4-\lambda_5 )+v_\chi (M_1+12 M_2) +\lambda_5 v_\phi^2\right]\,,
\end{align}

The couplings of a pair of Goldstone bosons to the $Z$ or $W$ are given by $i e C_{V s_1 s_2} \left(p_1 - p_2 \right)_{\mu}$, where $p_1$ ($p_2$) is the incoming momentum of incoming scalar $s_1$ ($s_2$) and the coefficients are
\begin{eqnarray}
C_{ZG^+G^{+*}} 
	&=& \frac{1}{2 s_Wc_W}\left(1 - 2s_W^2\right),  \\
	C_{W^- G^0 G^+} &=& \frac{i}{2 s_W}.
\end{eqnarray}

The couplings of a Goldstone boson and a physical scalar to a single vector boson are given by $i e C_{V s_1 s_2} \left(p_1 - p_2\right)_\mu$,
where $p_1$ ($p_2$) is the incoming momentum of incoming scalar $s_1$ ($s_2$).  The coefficients used here are given by
\begin{eqnarray}
	C_{Z H_5^0 G^0} &=& -2 i \sqrt{\frac{2}{3}}\frac{1}{s_Wc_W} \frac{v_\chi}{v}, \\
	C_{Z H_5^+G^{+*}} &=& C_{Z H_5^{+*} G^+} = -\frac{\sqrt{2}}{s_Wc_W}\frac{v_\chi}{v}, \\
	C_{W^+ H_5^0 G^{+*}} &=& -C_{W^- H_5^0 G^+} = \sqrt{\frac{2}{3}}\frac{1}{s_W} \frac{v_\chi}{v}, \\
	C_{W^+ G^0 H_5^{+*}} &=& C_{W^- G^0 H_5^+} = \frac{i \sqrt{2}}{s_W} \frac{v_{\chi}}{v}, \\
	C_{W^+ G^+ H_5^{++*}} &=& -C_{W^- G^{+*} H_5^{++}} = \frac{2}{s_W} \frac{v_{\chi}}{v}, \\
	C_{W^+ h G^{+*}} &=& -C_{W^- h G^+} = -\frac{1}{6 s_W v} \left(8 \sqrt{3} s_{\alpha} v_{\chi} - 3 c_{\alpha} v_{\phi} \right), \\
	C_{W^+ H G^{+*}} &=& -C_{W^- H G^+} = \frac{1}{6 s_W v} \left(8 \sqrt{3} c_{\alpha} v_{\chi} + 3 s_{\alpha} v_{\phi} \right).
\end{eqnarray}

The couplings of a Goldstone boson to two vector bosons are given by $i e^2 C_{s V_1 V_2} g_{\mu\nu}$, with
\begin{eqnarray}
	C_{G^+W^-\gamma} &=& C_{G^- W^+ \gamma} = \frac{v}{2 s_W}, \\
	C_{G^+W^-Z} &=& C_{G^- W^+ Z} = -\frac{v}{2 c_W}.
\end{eqnarray}

\subsection{Couplings involving ghosts}

Our calculation of the vector loop diagrams in the 't~Hooft-Feynman gauge requires the calculation of diagrams involving ghosts.  This enters only in the decay $H_5^+ \to W^+ \gamma$. 
The relevant term in the ghost Lagrangian involving an incoming $H_5^+$, incoming ghost $c^-$ and outgoing ghost $c^Z$ is
\begin{equation}
	\mathcal{L} \supset - \xi v_{\chi} \frac{e^2}{\sqrt{2} s^2_W c_W} \bar c^Z H_5^+ c^-,
\end{equation}
where $\xi = 1$ is the gauge-fixing parameter for the 't~Hooft-Feynman gauge.  The resulting Feynman rule for the Higgs-ghost-ghost vertex is $i e^2 C_{H_i c^- c^Z}$ with
\begin{equation}
	C_{H_5^+ c^- c^Z} = - \frac{v_{\chi}}{\sqrt{2} s_W^2 c_W}.
\end{equation}
In our conventions, the Feynman rules for a pair of ghosts coupling to a vector boson are 
\begin{eqnarray}
	\bar c^-(-k) c^Z W^-_{\nu}: && \quad i g c_W k_{\nu}, \\
	\bar c^-(-k) c^- \gamma_{\nu}: && \quad -i e k_{\nu},
\end{eqnarray}
where $-k$ is the incoming momentum of the incoming antighost; i.e., $k$ is the outgoing momentum of the outgoing ghost.

\section{LoopTools conventions}
\label{app:looptools}

We summarize here the conventions used by the LoopTools package~\cite{Hahn:1998yk} for the one-loop integrals that we have used in this paper.  The three-point integral for a diagram with incoming external momenta $p_1$, $p_2$, and $p_{12} = -p_1 - p_2$ and internal masses $m_1$, $m_2$, and $m_3$ is defined as
\begin{equation}
	\frac{i}{16 \pi^2} C_{0,\mu, \mu\nu}(p_1^2, p_2^2, p_{12}^2; m_1^2, m_2^2, m_3^2)
	= \int \frac{d^D q}{(2 \pi)^D} \frac{1, q_{\mu}, q_{\mu} q_{\nu}}
	{[q^2 - m_1^2] [(q + k_1)^2 - m_2^2] [(q + k_2)^2 - m_3^2]},
\end{equation}
where $k_1 = p_1$ and $k_2 = p_1 + p_2 = -p_{12}$.

The vector and tensor three-point integrals are decomposed into scalar coefficients according to
\begin{eqnarray}
	C_{\mu} &=& k_{1\mu} C_1 + k_{2\mu} C_2, \\
	C_{\mu\nu} &=& g_{\mu\nu} C_{00} + k_{1\mu} k_{1\nu} C_{11} + k_{1\mu} k_{2 \nu} C_{12}
	+ k_{2\mu} k_{1\nu} C_{21} + k_{2\mu} k_{2\nu} C_{22},
\end{eqnarray}
where $C_{21} = C_{12}$ due to the symmetry of $C_{\mu\nu}$ under permutation of Lorentz indices.  For compactness, when a sum of three-point integrals with a common set of arguments appears, we have specified the arguments only once at the end of the sum.

\section{Details of calculations in 't~Hooft-Feynman gauge}
\label{app:tHF}

In this appendix we give some details of the calculations in the 't~Hooft-Feynman gauge of processes that involve Goldstone bosons or ghosts.  This is relevant for the vector-scalar-scalar, scalar-vector-vector, and vector loop diagrams.

\subsection{Vector-scalar-scalar loop diagram}

In 't~Hooft-Feynman gauge there are two diagrams that contribute to this amplitude: one as shown by the third diagram of Fig.~\ref{fig:feyndiagrams} and one with the vector boson $X_1$ replaced by the corresponding Goldstone boson $G$.  The calculation of this second diagram is identical to the calculation of the scalar loop diagram [see Eq.~(\ref{eq:AsssLT})].  We write the contribution to the amplitude from these two diagrams as
\begin{equation}
	A_{Xss}^{H_i V \gamma} = S_{Xss} + S_{Gss},
\end{equation}
where
\begin{eqnarray}
	S_{Xss} &=& - \alpha_{\rm em}^2 Q_{s_2} C_{X_1^* H_i s_2} C_{s_2^* X_1 V^*} 
	\left[ -2 (C_{12} + C_{22} + 2 C_1 + 3 C_2 + 2 C_0) \right]
	(k^2, q^2, m_{H_i}^2; M_{X_1}^2, m_{s_2}^2, m_{s_2}^2), 
	\label{eq:SXss} \\
	S_{Gss} &=& - \frac{\alpha_{\rm em} Q_{s_2}}{\pi} C_{H_i G^* s_2} C_{V^* G s_2^*}
	\left[ C_{12} + C_{22} + C_2 \right]
	(k^2, q^2, m_{H_i}^2; M_{X_1}^2, m_{s_2}^2, m_{s_2}^2).
\end{eqnarray}

To combine these into a single expression, we examine the Goldstone boson couplings for the actual combinations of parent and internal particles in the decays of interest.  The scalar $s_2$ is always an  $H_5$ of nonzero electric charge, and the couplings of two $H_5$ states to a Goldstone boson are zero by custodial symmetry; thus the $S_{Gss}$ term contributes only to $H_3^+ \to W^+ \gamma$, not to $H_5^+ \to W^+ \gamma$ or $H_5^0 \to Z \gamma$.  Substituting the appropriate Goldstone boson couplings from Appendix~\ref{app:goldstonecoups}, $A_{Xss}^{H_i V \gamma}$ reduces to the expression given in Eq.~(\ref{eq:A_Xss}).  Note that the second line in Eq.~(\ref{eq:A_Xss}) contains a factor of $(m_{H_i}^2 - m_{s_2}^2)$, which is zero when $H_i$ and $s_2$ are both $H_5$ states.

\subsection{Scalar-vector-vector loop diagram}

In 't~Hooft-Feynman gauge there are four diagrams that contribute to this amplitude: one as shown by the fourth diagram of Fig.~\ref{fig:feyndiagrams}, two in which one of the gauge bosons $X_2$ in the loop has been replaced by the corresponding Goldstone boson $G$, and one in which both gauge bosons $X_2$ in the loop are replaced by Goldstone bosons $G$.  The calculation of the last of these diagrams is identical to that of the scalar loop diagram [see Eq.~(\ref{eq:AsssLT})].  We write the contribution to the amplitude from these four diagrams as
\begin{equation}
	A_{sXX}^{H_i V \gamma} = S_{sXX} + S_{sGX} + S_{sXG} + S_{sGG},
\end{equation}
where the subscripts denote the particles in the loop proceeding clockwise from the $H_i$ vertex.  The diagram corresponding to $S_{sXG}$ does not contribute to the $k^{\mu} q^{\nu}$ term in the amplitude, so $S_{sXG} = 0$.  The remaining amplitudes are
\begin{eqnarray}
	S_{sXX} &=& \alpha_{\rm em}^2 Q_{X_2} C_{X_2 H_i s_1^*} C_{s_1 X_2^* V^*}
	\left[ -C_{12} - C_{22} + 4 C_1 + C_2 \right]
	(k^2, q^2, m_{H_i}^2; m_{s_1}^2, M_{X_2}^2, M_{X_2}^2), 
	\label{eq:SsXX} \\
	S_{sGX} &=& - \alpha_{\rm em}^2 C_{X_2 H_i s_1^*} C_{V^* s_1 G^*} C_{G X^* \gamma}
	\left[ 2 C_{12} + 2 C_{22} - 2 C_2 \right]
	(k^2, q^2, m_{H_i}^2; m_{s_1}^2, M_{X_2}^2, M_{X_2}^2), 
	\label{eq:SsGX} \\
	S_{sGG} &=& - \frac{\alpha_{\rm em} Q_G}{\pi} C_{H_i s_1^* G} C_{V^* s_1 G^*}
	\left[ C_{12} + C_{22} + C_2 \right]
	(k^2, q^2, m_{H_i}^2; m_{s_1}^2, M_{X_2}^2, M_{X_2}^2).
\end{eqnarray}

To combine these into a single expression, we again examine the Goldstone boson couplings for the actual combinations of parent and internal particles in the decays of interest.  For $H_5^+ \to W^+ \gamma$ and $H_5^0 \to Z \gamma$, $s_1$ is always an $H_5$ state and $X_2$ is always a $W$ boson (of either charge).  Because the coupling of two $H_5$ states to a Goldstone boson is zero by custodial symmetry, $S_{sGG}$ does not contribute to these decays.  The remaining pieces are easy to combine using the relations between the Goldstone couplings and the corresponding gauge boson couplings, yielding the first line of Eq.~(\ref{eq:AsXX}) [note that the second line does not contribute for an initial-state $H_5$ because $s_1$ is also an $H_5$ state and hence $(m_{H_i}^2 - m_{s_1}^2) = 0$].

For $H_3^+ \to W^+ \gamma$, the situation is more complicated.  $s_1$ can be $h$, $H$, $H_5^0$, or $H_5^{++}$, and in all cases $S_{sGG}$ is nonzero.  For either of the $H_5$ states in the loop, the combination is again fairly straightforward and yields the expression in Eq.~(\ref{eq:AsXX}), with $S_{sGG}$ giving rise to the terms in the second line.  For $h$ or $H$ in the loop, the combination of $S_{sXX}$ and $S_{sGX}$ is again straightforward, yielding the first line in Eq.~(\ref{eq:AsXX}); the simplification of $S_{sGG}$ is non-obvious because of the complicated form of the $H_3^+ h G^-$ and $H_3^+ H G^-$ couplings, but it can be verified numerically that it also reduces to the terms in the second line of Eq.~(\ref{eq:AsXX}) for each diagram individually.

\subsection{Vector loop diagram}

In 't~Hooft-Feynman gauge, the last diagram in Fig.~\ref{fig:feyndiagrams} and its Goldstone boson substitutions do not contribute to the $k^{\mu} q^{\nu}$ term, so we only need to worry about the fifth diagram and its Goldstone and ghost substitutions.  There are nine diagrams: one as shown by the fifth diagram in Fig.~\ref{fig:feyndiagrams}, three in which a single gauge boson in the loop is replaced by the corresponding Goldstone boson, three in which two of the gauge bosons in the loop are replaced by their corresponding Goldstone bosons, one in which all three gauge bosons in the loop are replaced by Goldstone bosons, and a diagram with ghosts in the loop.  

We write the contribution to the amplitude from these nine diagrams as
\begin{equation}
	A_{X_1X_2X_2}^{H_i V \gamma} = S_{XXX} + S_{GXX} + S_{XGX} + S_{XXG}
	+ S_{GGX} + S_{XGG} + S_{GXG} + S_{GGG} + S_{\rm ghost},
\end{equation}
where again the subscripts denote the particles in the loop proceeding clockwise from the $H_i$ vertex.  The diagrams corresponding to $S_{XGX}$ and $S_{GXG}$ do not contribute to the $k^{\mu} q^{\nu}$ term in the amplitude, so they are zero.  Four more of the amplitudes can be read off from the scalar loop diagram [Eq.~(\ref{eq:AsssLT})] and diagrams computed earlier in this appendix [Eqs.~(\ref{eq:SXss}), (\ref{eq:SsXX}), and (\ref{eq:SsGX}), respectively]:
\begin{eqnarray}
	S_{GGG} &=& - \frac{\alpha_{\rm em} Q_{G_2}}{\pi} C_{H_i G_1^* G_2} C_{V^* G_1 G_2^*}
	\left[ C_{12} + C_{22} + C_2 \right] (k^2, q^2, m_{H_i}^2; M_{X_1}^2, M_{X_2}^2, M_{X_2}^2), \\
	S_{XGG} &=& -\alpha_{\rm em}^2 Q_{G_2} C_{X_1^* H_i G_2} C_{G_2^* X_1 V^*}
	\left[ -2(C_{12} + C_{22} + 2 C_1 + 3 C_2 + 2 C_0) \right]
	(k^2, q^2, m_{H_i}^2; M_{X_1}^2, M_{X_2}^2, M_{X_2}^2), \\
	S_{GXX} &=& \alpha_{\rm em}^2 Q_{X_2} C_{X_2 H_i G_1^*} C_{G_1 X_2^* V^*}
	\left[ -C_{12} - C_{22} + 4 C_1 + C_2 \right]
	(k^2, q^2, m_{H_i}^2; M_{X_1}^2, M_{X_2}^2, M_{X_2}^2), \\
	S_{GGX} &=& - \alpha_{\rm em}^2 C_{X_2 H_i G_1^*} C_{V^* G_1 G_2^*} C_{G_2 X_2^* \gamma} 
	\left[ 2 C_{12} + 2 C_{22} - 2 C_2 \right] 
	(k^2, q^2, m_{H_i}^2; M_{X_1}^2, M_{X_2}^2, M_{X_2}^2).
\end{eqnarray}

For the remaining diagrams, we specialize to the actual process of interest, $H_5^+ \to W^+ \gamma$, with $X_1 = Z$ and $X_2 = W^-$.  We can then use the explicit expressions for the triple-gauge and ghost vertices.  We obtain,
\begin{eqnarray}
	S_{XXG} &=& \alpha_{\rm em}^2 C_{Z H_i G^{+*}} C_{G^+ W^- \gamma} \frac{c_W}{s_W}
	\left[ C_{12} + C_{22} - 2 C_1 + 3 C_2 + 2 C_0 \right]
	(k^2, q^2, m_{H_i}^2; M_{X_1}^2, M_{X_2}^2, M_{X_2}^2), \\
	S_{XXX} &=& - \alpha_{\rm em}^2 Q_{W^-} C_{H_i W^- Z} \frac{c_W}{s_W}
	\left[ 10 C_{12} + 10 C_{22} + C_1 + 10 C_2 + 5 C_0 \right]
	(k^2, q^2, m_{H_i}^2; M_{X_1}^2, M_{X_2}^2, M_{X_2}^2), \\
	S_{\rm ghost} &=& 2 \alpha_{\rm em}^2 C_{H_i c^- c^Z} \frac{c_W}{s_W}
	\left[ C_{12} + C_{22} + C_2 \right] (k^2, q^2, m_{H_i}^2; M_{X_1}^2, M_{X_2}^2, M_{X_2}^2).
\end{eqnarray}
The last expression for $S_{\rm ghost}$ includes the contributions of the two ghost diagrams: one with $c^Z, c^-, c^-$ proceeding clockwise around the loop from the $H_5^+$ vertex, and one with $c^+, c^+, c^Z$ proceeding counterclockwise around the loop from the $H_5^+$ vertex (these are distinct because the antiparticle of the ghost $c^-$ is $\bar c^-$, not $c^+$).  These two diagrams give identical contributions.  

Inserting explicit expressions for all the couplings and combining all the terms is then relatively straightforward, and yields the expression for $A_{ZWW}^{H_5^+ W \gamma}$ given in Eq.~(\ref{eq:AZWW}).


\end{document}